**Statistical characterization of residual noise in the low-rank approximation filter framework, general theory and application to hyperpolarized tracer spectroscopy.**

R. Francischello[§], M.F. Santarelli[$], A. Flori[&], L. Menichetti[$], M. Geppi[#]

[§]Academic Radiology, Department of Translational Research and of New Surgical and Medical Technology, University of Pisa, Pisa, Italy
[&]U.O.C Bioengineering and Clinical Technology, Fondazione G. Monasterio CNR—Regione Toscana,56124 Pisa, Italy
[$] Institute of Clinical Physiology, National Research Council (IFC-CNR),56124 Pisa, Italy
[#] Dipartimento di Chimica e Chimica Industriale, Università di Pisa, via G. Moruzzi 13, 56124 Pisa (Italy)

Corresponding author: Roberto Francischello roberto.francischello@gmail.com

**Abstract:**
The use of low-rank approximation filters in the field of NMR is increasing due to their flexibility and effectiveness. Despite their ability to reduce the Mean Square Error between the processed signal and the true signal is well known, the statistical distribution of the residual noise is still undescribed.
In this article, we show that low-rank approximation filters are equivalent to linear filters, and we calculate the mean and the covariance matrix of the processed data. We also show how to use this knowledge to build a maximum likelihood estimator, and we test the estimator's performance with a Montecarlo simulation of a 13C pyruvate metabolic tracer.
While the article focuses on NMR spectroscopy experiment with hyperpolarized tracer, we also show that the results can be applied to tensorial data (e.g. using HOSVD) or 1D data (e.g. Cadzow filter).

**List of abbreviations:**
SVD: singular value decomposition
HOSVD: higher order singular value decomposition
FID: free induction decay
LI: linear indipendent

MLE: maximum likelihood estimator
LSE: least sqare
CRLB: Cramer Rao lower bound
NLLS: non-linear least square
SSA: singular spectrum analysis
AWGN: additive white Gaussian noise

# 1 Introduction

Low-rank approximation filtering is a noise reduction procedure based on the calculation of the best approximation of the original dataset under a rank constraint, which means that the new dataset is a linear combination of a fixed number of linearly independent components. Originally developed for matrix-shaped data, it is currently used for tensor-shaped data and vector-shaped data. The intuition behind low-rank filters is that usually signals are composed of a small number of Linear Independent (LI) components while noise is composed of many LI components, so they plan to exploit this difference to separate the signal from the noise and construct a filter.

Low-rank approximation filters are increasingly applied in the data analysis of a wide range of experiments, for instance in Nuclear Magnetic Resonance (NMR) spectroscopy, mass spectrometry[1], infrared spectroscopy[2], fluid dynamics[3], hyperspectral imaging[4], and biomedical imaging[5]. Their use is a well-documented practice in both NMR and MRI and in particular in the fields of hyperpolarized NMR and diffusion-weighted MRI[6].

Historically, the idea of low-rank approximation in NMR started as a quantification tool. In the second half of the 90's Stoyanova et al.[7] proposed the use of PCA to quantify the variation of a single peak area in a series of multiple spectra. This method was further refined to exploit complex values input data[8] and to consider the effect of line shape variation across multiple samples[9]. The use of PCA has as a secondary effect a significant reduction in the signal's noise and therefore an increase in the accuracy obtained when determining peak areas. While these papers did not directly describe the use of low-rank approximation as a noise reduction technique, they provided useful insight into the use of Singular Value Decomposition (SVD) on NMR signals.

Very recently, low-rank approximation filters were extensively applied to process both matrix and tensor datasets in different NMR fields like studies of hyperpolarized contrast agents[10–22], Magnetic Resonance Spectroscopy Imaging[23–32] (MRSI), diffusion imaging[32–59], functional MRI[60–70], CEST-MRI[71–83], $T_2$ or $T_1$ mapping[43,45,84–90], NMR spectroscopy[22,91–99], and simple two or three-dimensional $T_2$ or $T_1$ weighted images[100–103].

Low-rank approximation filters are data-driven algorithms that require minimum human intervention, the only user-selected parameter being the rank of the approximating matrix. This parameter determines the stiffness of the filter because the rank of the reconstructed signal is equal to the number of linearly independent components in the reconstructed signal. There is a trade-off in choosing the number of singular values to use for the low-rank approximation; using too few singular values may cause a loss of information because not enough LI components are present to reconstruct exactly the original signal but suppress more noise; using too many singular values limits the distortion of the true signal but also preserve more noise. Many methods for the automatic selection of the optimal rank selection have been proposed, mainly based on the random matrix theory[54,104,105] or the residual error in the singular vector[3,106]. Among those based on the random matrix theory, there is an important subclass of methods devoted to the estimation of an optimal threshold in the intensity of the singular value that minimizes the mean square error between the noiseless unknown signal and the reconstructed matrix[104,105], i.e. the denoised measurement matrix. These methods are extremely important because they also provide an intrinsic theoretical justification for the use of low-rank approximation filters even if the noiseless data do not have an intrinsic low-rank structure. Indeed, the applications of these methods proved that under weak conditions (e.g. the presence of additive gaussian noise and an element-wise stochastic perturbation of the true signal) a reconstruction rank exists that minimizes the mean square error between the true signal and the low-rank approximation of the measurements. Leeb[107] described an extension to the classical low-rank approximation filter based on the manipulation of the intensity of the singular values to directly minimize the mean square error between the reconstructed matrix and the pure signal instead of the minimization of the differences of the reconstructed matrix with the measurement matrix. This required the definition of a shrinking function, a multiplicative weight applied to the singular values, based on their intensity and the noise statistic. The new singular values obtained after the application of the shrinking function are used to form the reconstructed matrix following the standard procedure for the low-rank approximation problem.

Despite the presence in the literature of all of these studies on the possible applications of the low-rank filter family, to the best of our knowledge, only one study[24] describes the effects of these filters on the statistical distribution of the residual noise and the shape of the signal. The term "residual noise" is here used to describe the components that are not the NMR signal and are present in the measurement after the use of a low-rank approximation filter. NMR is an intrinsically quantitative method, therefore the acquired signal is commonly analyzed with techniques preserving quantitatively, such as model fitting.

Clark et Chiew[24] provide an estimation of the covariance matrix of the filtered signal for a two-dimentional dataset with Additive Gaussian Wight Noise (AGWN) using the results form Chen et al.[108] and Song et al.[109].

Knowing the statistic of the dataset distribution is important to provide not only accurate and precise estimation but also to understand the statistical properties of the estimated parameters. Some authors[22,41,61] pointed out that the residual noise is correlated and raised concerns about the impossibility of using advanced analysis methods[110–116] on filtered signals due to such a correlation. In this paper, we will show that this concern is well-grounded because the low-rank approximation filter can be seen as a non-local mean filter that introduces correlations between different portions of the dataset. The presence of correlation between different portions of the data has for instance an important impact on image analyses because it can introduce long-range correlation between different spatial regions. This problem is not confined to the image datasets: multidimensional NMR datasets, such as in-vivo reaction monitoring using a hyperpolarized contrast agent, can be analyzed through multidimensional fits[110] to increase the accuracy of the estimation. The standard implementation of this method assumes that the noise is uncorrelated; the correlation, introduced by the low-rank approximation filter, is in contrast with this assumption and reduces the estimator performance. However, if the data correlation matrix is known it is possible to use this additional knowledge to increase the estimator performance.

## 2 Theory and Calculation

In this section, we calculate the effect of a low-rank approximation filter based on SVD on NMR signals. We will derive these results for a simple TSVD decomposition filter and then we will generalise it to a multidimensional dataset, one-dimensional dataset and filter based on singular value shrinking.

*2.1 TSVD filters in a 2D signal are linear projections*

Let **S** be our signal matrix: **S** is a *m* x *n* complex value matrix in which every row is an ideal FID collected at a different time during the experiment without any random noise. The rank of **S** depends on the number of independent components that are in the signal for example the number of metabolites observed and depends on their spectral and metabolism kinetics. Without loss of generality, we assume that the rank of **S** is k.

Let **N** be a complex *m* x *n* matrix that represents the noise of the measure. This matrix has full rank. Finally, **M=S+N** represents the measured experimental data and has full rank.

The construction of a low-rank filter requires the solution of the following optimization problem:
$$\min_{X}|M - X|_F \quad rank(X) = r$$
Where **X** is the filtered signal, and $r$ is the rank we choose to use for the approximation. The choice of $r$ could be made based on empirical consideration or theoretical reason depending on the properties of **M** and our knowledge of the system. The Eckart–Young–Mirsky theorem shows that the solution to this optimization problem could be calculated using the TSVD.

Using the SVD we can write $M=U\Sigma V^*$. Let $\Sigma_1$ be the diagonal matrix whose elements $\sigma_{1\,ii} = \sigma_{ii}$ for $1\leq i \leq r$ and zero otherwise. Then $M_1 = U\Sigma_1 V^*$ is equal to **X** and it is the solution to the low-rank approximation filter. This procedure is called TSVD because the singular values were truncated in the sense of put to zero.

While the SVD is a non-linear function of its input, once the matrix $M_1$ is known it is easy to calculate a matrix $P_1$ such that its multiplication with the original signal matrix **M** produces the filtered matrix $M_1$. The matrix $P_1$ is equal to the product of the first r rows of the matrix **U**, $U_1$, with their conjugate transpose $U_1^*$. This is a projection matrix on the subspace spanned by the first r rows of matrix **U**. The demonstration is trivial due to the orthonormality of the rows of matrix **U**. We can generalize this result to the singular values shrinking procedure if we use a generic projection $A=UHU^*$ with **H** a diagonal matrix with elements equal to the shrinking function used for shrinking the singular values such as the $i^{th}$ elements of the diagonal is equal to the shrinking factor of the $i^{th}$ singular values. The truncated case is obtained if **H** is a diagonal matrix with only the first r elements of the diagonal different from zero.

*2.2 Statistical properties of TSVD filter on 2D data*

The studies of the statistical properties of the signal matrix **M** are easier if the matrix is reshaped into a 1D vector, **b**, and studied in terms of its var-covariance matrix **C**.

To study the effect of the filter on the measured signal statistical distribution, we need to calculate the projection matrix **T**, having dimensions *mn x mn*, that sends the vector **b**=vect(**M**) into the vector $b_1$=vect($M_1$)= vect($P_1M$), Using row-major order vectorization properties[a], we can write:

$$vect(P_1MI) = (P_1 \otimes I)vect(M) = Tb$$

Where $\otimes$ is the Kronecker product and **T** is the matrix $(P_1 \otimes I)$.

Since we have shown that the low-rank filter is a linear transformation, we can derive some general results on the mean and the covariance matrix of the transformed dataset.

The mean operator is linear; therefore, the mean of the filtered data, $\mu_1$, is equal to **T μ**.

The covariance operator is bilinear, therefore the transformed covariance matrix, $C_1$, is equal to $TCT^*$ where **C** is the covariance matrix before the transformation.

These two results, based only on the expectation value operator linearity, are valid for each noise distribution with a defined mean and covariance matrix; they do not depend in any way on the statistic noise distribution and describe only the mean and the covariance of the transformed dataset.

While useful, these results do not provide a complete characterization of the noise statistics after the use of low-rank filters. Nevertheless, if the noise is purely additive, the mean of the experimental data is equal to the sum of the pure signal and the noise mean. Therefore, knowing how the mean changes after the use of a low-rank filter, it is possible to quantify the distortion of the pure signal due to the data processing. Similarly, the knowledge of the transformed covariance matrix provides important insight into the data correlation which is useful for example for some statistical analysis frequently used in the field of fMRI.

---

[a] the choice between row-major order or column-major order is arbitrary, we choose the row-major vectorization because it is the standard one for numpy which is the library used for the numerical part of the paper

If the original covariance matrix is a multiple of the identity matrix, then $\mathbf{C_1} = \sigma^2(\mathbf{TT^*})$. Combining this result with the properties of the Kronecker product we obtain(proof in appendix A prop. 1):

$$\mathbf{C_1} = \sigma^2(\mathbf{P_1 P_1^*}) \otimes \mathbf{I}$$

$\mathbf{C_1}$ has a sparse form due to the Kronecker product with an identity matrix. Since $\mathbf{P_1}$ is a non-trivial projector, it is not invertible. Then $\mathbf{P_1} \mathbf{P_1^*}$ is not invertible and therefore due to the Kronecker product properties $\mathbf{C_1}$ is not invertible as well.

For the remaining part of the article, we will assume that the noise is an Additive White Gaussian Noise, which is a good approximation for the noise in complex value NMR signal.

*2.3 Least Square estimator on filtered signals*

Recent publications show that the use of low-rank approximation for noise reduction produces an improvement in the quantification of experimental signals[13,91] using a Least Squares Estimator (LSE). Based on the results from the previous section we have two observations on why the use of LSE on low-rank filtered signal is suboptimal.
Observation 1: The filter alters the signal shape, therefore even if the fitting function properly describes the true signal shape and if the fitting parameters are equal to the true signal parameters there will be a mismatch between the curve fitting function and the filtered true signal due to the distortion introduced by the filter.
Observation 2: The LSE estimator is mathematically equivalent to the MLE for a deterministic signal with AWGN. If the noise is correlated then the statistical justification for the use of LSE is no longer valid.
None of these issues is in itself a sufficient condition for the parameters estimated by LSE method to be unreliable, nonetheless, it is reasonable to expect that there are conditions, for instance, the shape of the signal, for which the LSE estimator produces unsatisfactory results. It, therefore, turns out to be of interest the possibility of defining an MLE for the low-rank filtered signal.

*2.4 Maximum Likelihood Estimator*

We recall that for complex numbers, a dataset y, a model f(x), and a zero-mean correlated Gaussian noise the log-likelihood function is:

$$-[ln(|\mathbf{C}|) + (y - f(x))^* \mathbf{C}^{-1} (y - f(x)) + k\,ln(\pi)]$$

Where $\mathbf{C}$ is the covariance matrix of the correlated Gaussian noise. If the multivariate normal distribution has a singular covariance matrix, then the More-Penrose inverse should be used instead of its inverse $\mathbf{C}^{-1}$.
In section 2.2 we showed that the covariance matrix for the filtered signal is singular hand is equal to $\sigma^2(\mathbf{P_1 P_1^*}) \otimes \mathbf{I}$.
Once the constant terms are dropped the log-likelihood function takes the form of:

$$-\left(\mathbf{T}(b - f(\theta))\right)^* \mathbf{C}^+ \left(\mathbf{T}(b - f(\theta))\right)$$

Where we projected both the signal and the model onto the sub-space generated by the low-rank approximation filter, f() is the model and θ is the model's parameters vector, and $\mathbf{T}$ is equal to $\mathbf{P_1} \otimes \mathbf{I}$ is the projection generated by the low-rank approximation filter.
This loglikelihood function could be rewritten in dense matrix form to increase the numerical efficiency obtained (proof in Appendix A prop. 2):

$$-1/\sigma^2 \|\mathbf{DA}\|_F$$

Where **D** is equal to $P_1^+P_1$, **A** is the matrix of the differences between the experimental data **M** and $F(\theta)$, with $F(\theta)$ being the spectral-metabolic model in matrix form, and $\| \ \|_F$ is the Frobenius norm. If $P_1$ is an orthogonal projection, for example for a low-rank approximation filter based on TSVD, then (proof in Appendix A prop. 3) $P_1^+P_1 = P_1$.

The MLE for low-rank approximation filtered signal with additive zero-mean white Gaussian noise is obtained by projecting the difference between the fitting function and the experimental data.

*3.5 Generalization*

All the results reported in this section hold for any linear transform of our data independently of its origin. Therefore, these results also apply to any low-rank approximation filter based on the use of SVD on datasets that can be represented more naturally as tensors or vectors.

The use of a low-rank approximation filter in three or more dimensions involves the solution of the tensor low-rank approximation problem, which is far more complex than the matrix approximation problem[117]. Indeed, even the generalization of the concept of matrix rank to a tensor is not trivial, therefore we refer the interested reader to the specialistic literature[117,118].

The Higher Order Singular Value Decomposition[119] (HOSVD) is one of the possible generalizations of the SVD that can be used to find a low-rank approximation of a tensor. This decomposition is obtained with an iterative procedure that involves many reshaping of the tensor and multiple applications of the SVD. The full details are provided in the supplementary material, here we just note that the final results of the low-rank approximation obtained using the HOSVD is equivalent to a series of permutation of the flattened tensor alternated with orthogonal projection in a similar way to the SVD case. This procedure produces a quasi-optimal approximation of the original tensor. Since all the involved operations are linear, the results from this section can be applied to the HOSVD low-rank approximations.

The vector-shaped dataset could be converted to a matrix before using the low-rank approximation filter. In the Cadzow filter[2,120–128] or the Singular Spectrum Analysis[129–131] (SSA) method, the vector is first transformed into a Henkel matrix and then a low-rank approximation of this matrix is calculated, and finally, the rank-reduced Hankel matrix is transformed back into a 1D vectors. Since all those operations are linear the result from this section could be used to calculate the residual noise statistics. The full calculation is provided in the supplementary material.

In the introduction, we briefly describe the practice of singular values shrinkage as a generalization of singular values truncation and its benefits. From a mathematical point of view, the singular values shrinking is equivalent to a linear transform of the measurement data with a matrix that is not an orthogonal projection. In the end, the optimization problem for the MLE estimator for the truncation or the shrinking of singular values is equal, see supplementary material for the calculation. Therefore, while providing better results in terms of MSE the singular values shrinkage produces results equivalent to the singular values truncation if the MLE is used.

# 3 Methods

NMR spectroscopy experiments with hyperpolarized $^{13}$C tracers could be used to investigate metabolic pathways. A hyperpolarized metabolic tracer, such as $^{13}$C-pyruvate, is injected into the subject and multiple FIDs are acquired in close succession, to observe the conversion of the tracer into its metabolite. The standard data analysis procedure[132,133] is to analyze each measured signal,

FID or spectrum, to obtain the concentration of the metabolites at each measurement time. Then, the concentration signal is analyzed with a metabolic model that describes part of the physio-chemical parameter that governs the metabolic flux e.g., the apparent metabolic reaction rate, the $T_1$ relaxation time, and the tissue input function. It is possible to combine the spectral and the metabolic model into a single spectral-metabolic model that describes the change of signal intensity both during each FID acquisition and between FID acquisitions. This spectral-metabolic model is fitted on all the spectra in the datasets at the same time.

Usually, this problem is solved with a non-linear least square fit of the model on the data, this estimator is equivalent to the Maximum Likelihood Estimator (MLE) only if the noise is AWGN. MLEs have several theoretical advantages over least square estimators, e.g., an MLE attains Cramer Rao Lower Bound (CRLB) in an asymptotical sense. The CRLB is a lower bound on the covariance matrix values of parameters predicted with an unbiased estimator. The MLE is calculated form the likelihood function of the data distribution.

*3.1 Noiseless Data generation*

For the numerical simulation, we used Python with Numpy(1.20.3)[134], Numba (0.53.1)[135], and Scipy(1.7.1)[136] libraries, the plots were made using Matplotlib[137].

We generated the synthetic dataset according to the model presented in[133,138]. Full information on the spectral and metabolic parameters used for the simulation is reported in the supplementary material, and the Python function that produces a noiseless dataset is available at reasonable request from the authors.

We are simulating the injection of hyperpolarized $^{13}$C-pyruvate into a living organism and its subsequent metabolization according to the model proposed in Zierhut et al.[138]. Every dataset is a collection of FIDs measured at a different time after injection, every 2s, during the conversion of $^{13}$C pyruvate into its metabolites: $^{13}$C-lactate, $^{13}$C-bicarbonate, and $^{13}$C-alanine (in the rest of the article we drop the $^{13}$C sign to simplify the notation). The FIDs are arranged into a matrix. In the first phase of the experiment, i.e. the first 6 FIDs, the pyruvate is injected into the subject, and we observe a sharp increase in the pyruvate signal. During the rest of the experiment, there is a loss of signal intensity, for each metabolite, due to $T_1$ relaxation and radio frequency excitation, and there is a transfer of signal intensity from the pyruvate to its metabolites due to the metabolic conversion of pyruvate.

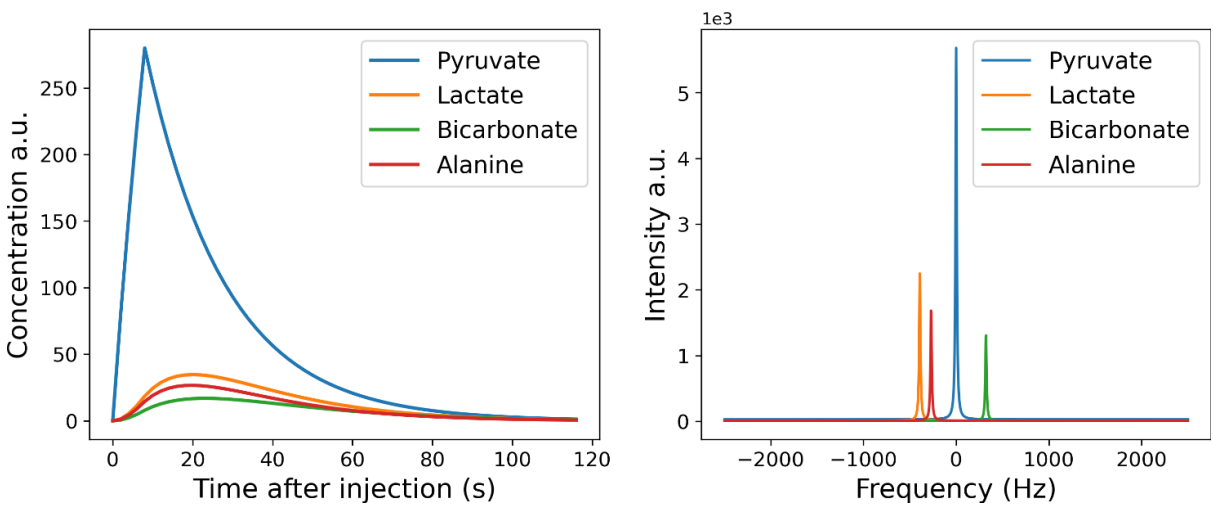

Figure 1: *The synthetic signal. Left: the metabolite signals intensity (y-axis) dependency on the time (x-axis). Right: the metabolite spectral components after 40s from the injections.*

The metabolite signal intensity at each time step follows the descending order: pyruvate, lactate, alanine, and bicarbonate. The exact curve profile for each metabolite and the signal spectrum after 40 seconds from the simulation of the tracer injection are reported in Figure 1.

*3.2 Model fitting*

We were interested in comparing the fitting performance of the standard data with that of the fitting of filtered data using two different models: the naïve nonlinear least square (without any correction for the effect of the low-rank approximation filter), and the proper ML estimator for the filtered data.
This information could be obtained by the Montecarlo simulation of the model fitting. We generated 500 datasets with added complex white Gaussian noise. Those datasets constitute a Montecarlo sample, and each sample is characterized by noise intensity and statistical distribution. The noise intensity is equal for all the FIDs in a dataset, i.e. it is constant during the whole hyperpolarization experiment. We measured the SNR as the ratio between the highest signal value in the datasets and the noise standard deviation. We considered SNR levels growing from 1 to 2048 as a power of 2 totalling 12 SNR levels. Therefore, we had 12 Montecarlo samples with a sample size of 500 datasets.
We obtained the filtered data using the SVD denoising and different numbers of singular values: 1, 2, 3, 4, 5, 6, 10, 20, 40, and 59. The filter built with 59 singular values should leave the signal unchanged and is used as a consistency check.
Then we fitted the spectral-metabolic model, with the same random starting points, on the different datasets using the scipy.optimize.minimize function with the Powell method. This is a good compromise between execution time and performance compared with the other methods available within the scipy library.

*3.3 Fisher information matrix and CRLB*

We wanted to calculate the Fisher Information (FI) matrix for our standard data model fitting and the filtered data ML estimator. The FI matrix is defined as:

$$FI(\theta)_{ij} = -E\left[\frac{\delta^2}{\delta\theta_i \delta\theta_j} \log f(y; \theta)\right]$$

We could obtain the expected value numerically by the Montecarlo simulation.
We generated a single dataset for each SNR level studied (from 1 to 2048 log spaced points, ten points for a decade), and defined the SVD filter with singular values numbers 1, 2, 3, 4, 5, 6, 10, 20, 40, 59. The simulation that uses all the singular values, 59, was used as an internal test for the correctness of the calculation.
Then, we created 100 new datasets with the noise sampled from the previously established noise distribution and applied the SVD filter matrix to them. By doing so we could study the effect of the SVD filter on the noise statistical distribution. Then, we defined the log-likelihood function for both standard data fitting and filtered data using the JAX library[139]. JAX provides functions for algorithmic differentiation and GPU acceleration, so we could easily calculate the Hessian of the log-likelihood in $\theta_0$ and obtain the FI matrix from their average.
It is important to execute the calculation into a 64-bit floating point: due to the numerical instability of some of the operations involved in the Hessian calculation the use of a 32-bit floating point could lead to wrong results. This error was enlightened by the comparison of the Hessian calculated using the 59 singular values filter and the raw data one: the two Hessians should be equal because the 59 singular values filter is just the original signal, but we found different values due to the rounding error in floating point arithmetic. Indeed, the different operations used to calculate the Hessian matrix for the two datasets, which should be equal, produced a different outcome.

Once the FI matrix was obtained, we aimed at calculating the CRLB. Unfortunately, the FI matrix has a high condition number. Therefore, the CRLB matrix obtained may be different from the true CRLB matrix. While we are aware of this limitation, we think that the calculated CRLB still provides significant information.

# 4 Results

Low-rank filters are increasingly used to process NMR data. Nevertheless, the statistical properties of residual noise are largely unexplored. This paper aims to characterize the effect of low-rank filters on residual noise both in the general case and in the case of additive Gaussian noise.
We proposed a new interpretation of the low-rank approximation filter as a linear projection on the pseudo signal subspace. From this interpretation we derived the statistical properties of the residual noise, first describing the mean and the covariance matrix for a generic noise distribution and then providing the likelihood function and the MLE for the spectral-metabolic fitting problem with additive white Gaussian noise.
All the graphics follow the same convention: the green star marker represents the results for the untreated data, and the solid lines represent the treated data. The solid lines' color shade encodes the number of singular values used in the definition of the low-rank approximation filter, the darker the shade the higher the number of singular values used.

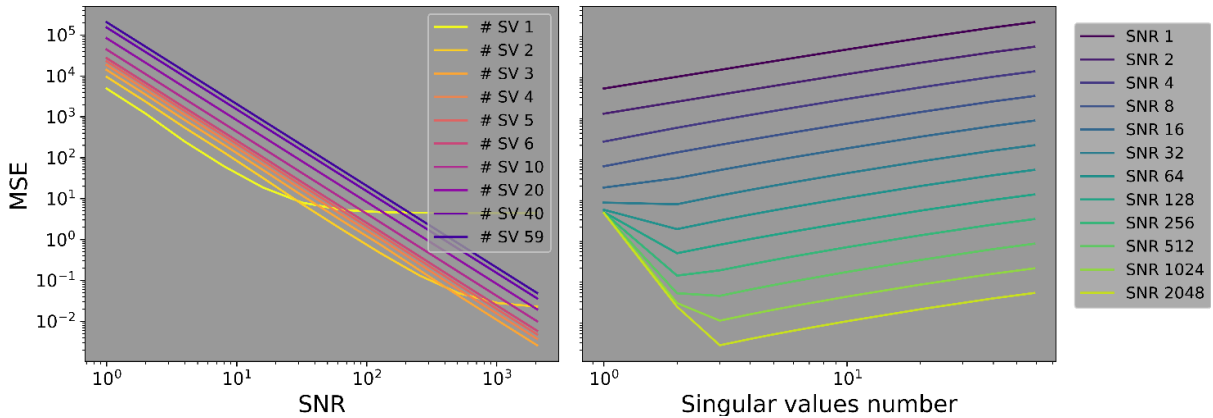

Figure 2: *Low-rank filters effect on Mean Square Error (MSE). Left: dependence of signal's MSE (y-axis, log scale) on the measurement SNR (x-axis, log scale); each curve corresponds to a different number of singular values used to build the low-rank filter, reported in the legend. Right: dependence of signal's MSE (y-axis, log scale) on the number of singular values used to define the filters (x-axis, log scale); each curve corresponds to a different measurements SNR, reported in the legend.*

In figure 2 the Mean Square Error (MSE) is reported *vs* both the SNR of the experimental signal and the number of singular values used to build the low-rank approximation filter. MSE is a direct measure of the filtered signal fidelity to the true signal and measures both the precision and the accuracy. Being a squared measure, MSE is sensible to the presence of outliers, and it does not consider the relative weight of the errors compared to the noise intensity. Finally, MSE is symmetric for positive and negative errors. Additional qualitative information on the effects of the low-rank approximation filter on this dataset is available in the supplementary material, supplementary figures from 21 to 26 describe the signal, its singular vectors, and singular value and how they change in the presence of different levels of noise, supplementary figure 27 reports the comparison of the noisy signal with the filtered signal for a selection of SNR levels and times after the injection.

The Montecarlo simulation produces high dimensional outputs: the LSE on the raw data produces a 12 (SNR level) x 500 (Montecarlo sample size) x17 (number of parameters) tensor while the estimators on the filtered data produce two 10 (filter strength) x 12 (SNR level) x 500 (Montecarlo sample size) x17 (number of parameters) tensor. The Root Mean Square Errors (RMSE) between the estimated parameter and its true value over the Montecarlo sample is used to summarize the estimator performance for each SNR level and filter strength if the estimator operates on filtered data.

Here we report and comment on only the results for the most significant parameters (from an applicative point of view), i.e. the apparent reaction constant for the pyruvate conversion in lactate ($k_{pl}$), and the apparent reaction constant for the pyruvate conversion in bicarbonate ($k_{pb}$).

The hyperpolarized pyruvate tracer allows the direct observation of the cell metabolism, the $k_{pl}$ is directly related to the lactate metabolism while $k_{pb}$ describes the aerobic metabolism. Together they are useful for studying both healthy and pathological states. The results for all parameters are reported in the supplementary figures from 1 to 16.

This choice highlights the estimators' performance dependency on the SNR levels and the filter strength measured as the reconstruction rank in the approximation problem.

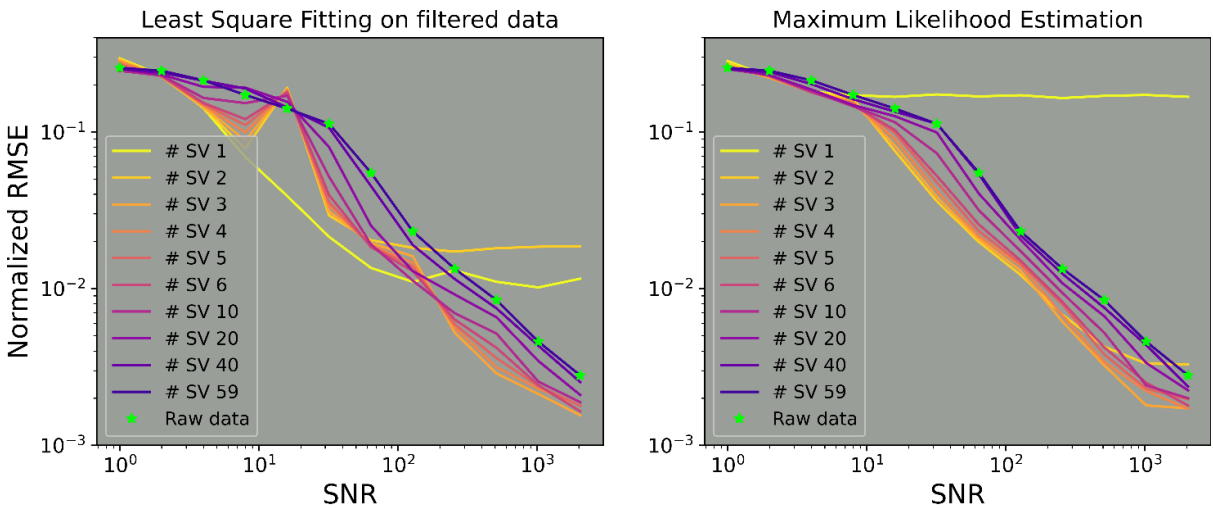

Figure 3: *Normalized Root Mean Square Error (nRMSE) of apparent Lactate kinetic constant. Left: nRMSE (y-axis, log scale) dependence on the SNR (x-axis, log scale); the least square estimators on the filtered and the original data are reported as solid lines and green stars, respectively. Right: nRMSE (y-axis, log scale) dependence on the SNR (x-axis, log scale); the solid lines and the green stars indicate the values for the maximum likelihood estimator on the filtered data and for the least square estimator on the original data, respectively.*

In figure 3 the trends of normalized RMSE (nRMSE) *vs* the SNR for the three estimators of $k_{pl}$ are reported.

The RMSE is normalized on the parameter value to highlight the effectiveness of the estimator, a value of 1 means that the error is of the same magnitude as the parameter a value of 0.01 means that the error is two orders of magnitude lower than the parameter.

The Maximum Likelihood Estimator on unfiltered data (MLE-RAW) is used as a reference for both the estimator on filtered data, the Least Square Estimator on the filtered data (LS-SVD), and the Maximum Likelihood Estimator on both filtered and unfiltered data (MLE-SVD).

Please note that while both the MLE-RAW and MLE-SVD estimators are based on the maximum likelihood framework their mathematical formulation is different because of the statistical differences due to the filter effects on the data

The use of low-rank approximation filters produces a significant reduction in RMSE at various SNRs compared to the standard estimation procedure on the unfiltered data. The low-rank filters are so effective that the RMSE of the $k_{pl}$ for the MLE-SVD estimator is lower (for all SNR levels higher than 16) than the RMSE of the standard estimator even when compared with a signal with a double SNR.

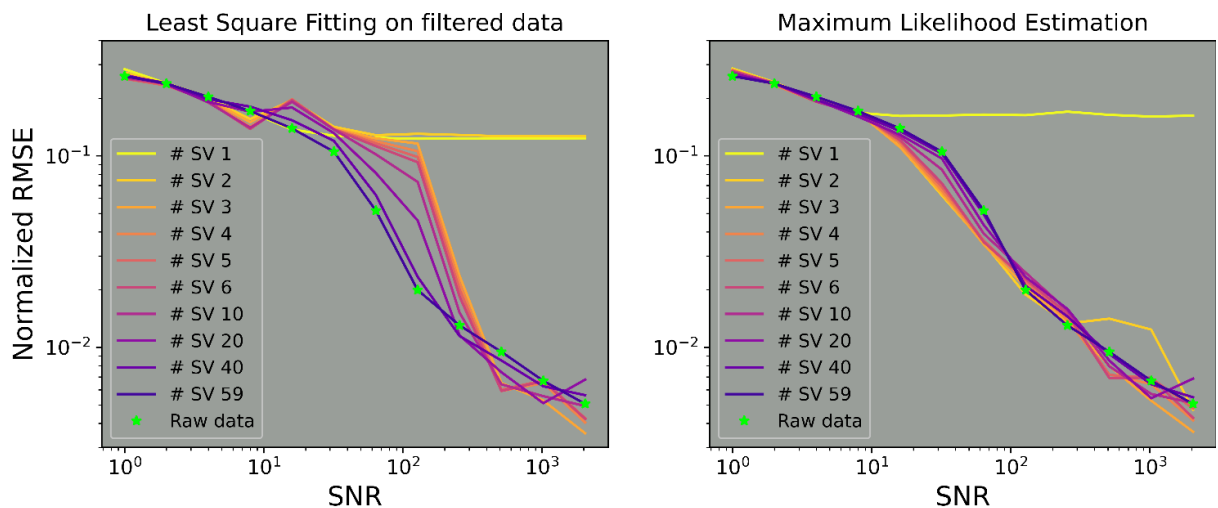

Figure 4: *nRMSE of apparent Bicarbonate kinetic constant. Left: nRMSE (y-axis, log scale) dependence on the SNR (x-axis, log scale); the least square estimators on the filtered and the original data are reported as solid lines and green stars, respectively. Right: nRMSE (y-axis, log scale) dependence on the SNR (x-axis, log scale); the solid lines and the green stars indicate the values for the maximum likelihood estimator on the filtered data and for the least square estimator on the original data, respectively.*

In analogy to what reported in figure 3 for $k_{pl}$, in figure 4 the trends of nRMSE *vs* the SNR for the three estimators of $k_{pb}$ are reported. These trends are qualitatively similar to those above described for $k_{pl}$.

In figure 5 the square roots of CRLB of the apparent lactate kinetic constant are reported vs the SNR level for the least square estimator on the raw data and for the MLE on the low-rank filtered data. The curves corresponding to the low-rank filter with 1 and 2 singular values are interrupted because the CRLB calculated for SNR levels take a negative value. CRLB should be positive, and this error is probably due to numerical errors in the inversion of the Fisher information matrix. This error could just be related to the use of finite precision (we observed a similar phenomenon when we tried to speed up the calculation using 32-bit floating-point precision instead of 64-bit) or the Fisher information matrix could be truly singular because we are trying to fit a four independent components model after we have projected it onto a subspace which is defined by 1 or 2 independent components. The model under study has 4 independent components, corresponding to the 4 metabolites, and we considered the low-rank filter constructed using less than 4 components to study how reliable this method is even if we strongly underestimate the number of linearly independent components of the signal. The complete reports on the CRLB estimation are reported in the supplementary figures form 17 to 20.

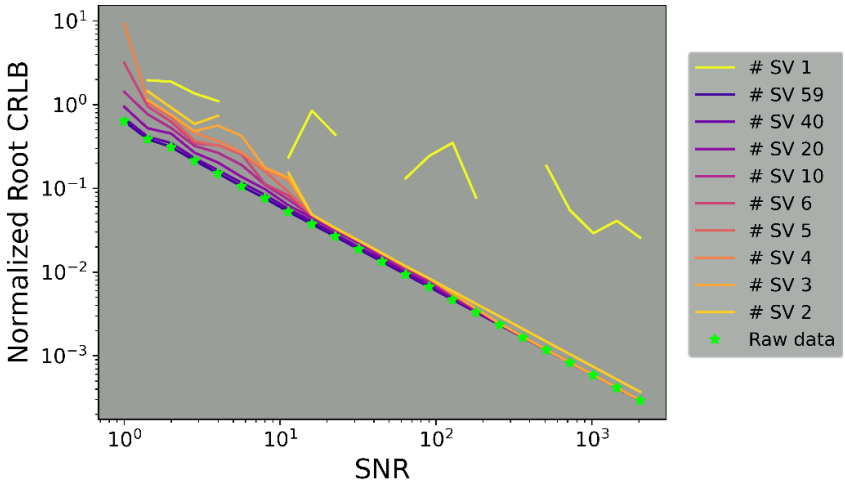

Figure 5: *Dependence of the normalized CRLB (y-axis, log scale) of apparent Lactate kinetic constant on the SNR (x-axis, log scale); the CRLB for the MLE on the filtered data and for the least square estimator on the original data are reported as solid lines and green stars, respectively.*

Finally, in figure 6 the trends of nRMSE *vs* the SNR for the three estimators of the effective bicarbonate spin-lattice relaxation rate are reported. This is a clear example of low-rank approximation failure. Both the LS-SVD and the MLE-SVD have a higher RMSE compared to the MLE on the raw data for various SNR levels. LS-SVD on filtered data shows a higher Normalized RMSE compared to MLE-SVD on almost all SNR levels considered. The low-rank filter improves the RMSE of both estimators, least square and maximum likelihood, at SNR equal to or higher than 1024.

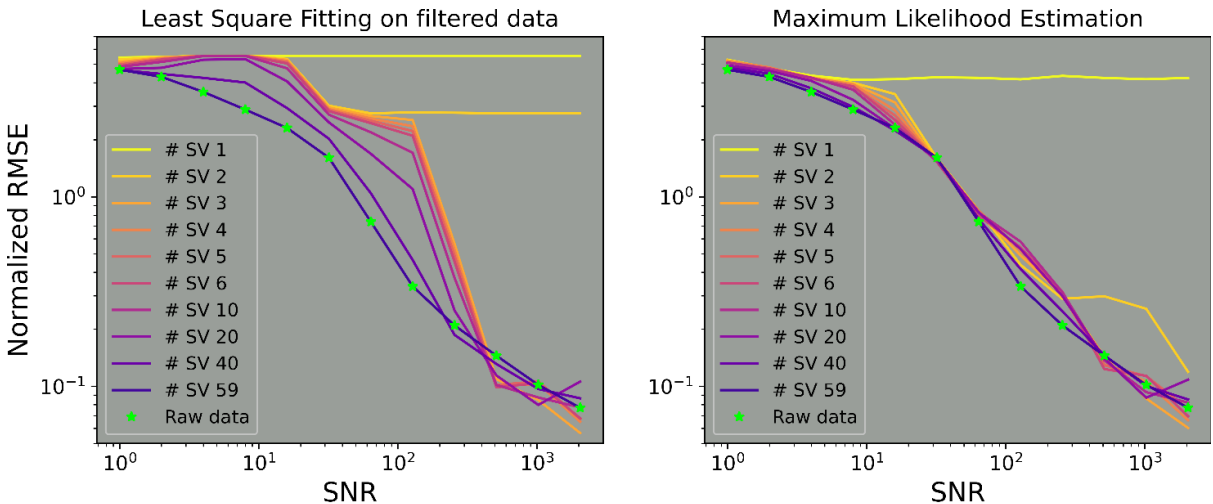

Figure 6: *nRMSE of effective Bicarbonate spin-lattice relaxation rate. Left: nRMSE (y-axis, log scale) dependence on the SNR (x-axis, log scale); the least square estimators on the filtered and the original data are reported as solid lines and green stars, respectively. Right: nRMSE (y-axis, log scale) dependence on the SNR (x-axis, log scale); the solid lines and the green stars indicate the values for the maximum likelihood estimator on the filtered data and for the least square estimator on the original data, respectively.*

# 5 Discussion

The main result of the paper is the description of a linear filter that provides the same results as the low-rank approximation filter, leading to: a) the general characterization of the filtered data mean and covariance matrix, b) to the definition of the residual noise distribution in the case of additive Gaussian noise c) the definition and characterization of a MLE for signal with AGWN.
The linear transform used to filter the data is rank deficit and not invertible, and the lack of invertibility has an important consequence on the MLE: indeed, parameters obtained with MLE are invariant under the application of an invertible transform of the experimental data. Therefore, the non-invertibility of the transform provides the opportunity for a variation in the estimated parameters and a potential reduction in their estimation error.
The previous results from Clark et Chiew[24] originate from the solution of the matrix completion problem. The aim of this method is to estimate a signal from a noise-corrupted undersampled measurement, therefore it is a generalization of the low-rank approximation problem where the signal is sampled in all interesting points. Chen et al.[108] proposed a procedure to approximate the uncertainty in the recovered signal, in our terminology the residual noise, assuming that the noise is AWGN and the SVD of the true signal is known. On the other hand, Song et al.[109] proposed to approximate the SVD of the true signal with the SVD of the denoised signal because, if the low-rank filter performs well, the filtered signal should be similar to the noiseless signal. Finally, our estimator differs from the already published ones from a theoretical point of view.
In the Chen et al. and derivated works the focus is on the solution of an optimization problem to solve the low-rank approximation problem and the studies of the solution statistical distribution. In our work, we use the solution of a low-rank approximation problem to build a linear filter that produces the same numerical output as the low-rank approximation problem. In this context, we consider the coefficients of our filter, the elements of the projection matrix $\mathbf{P_1}$, as a deterministic number and not a random variable. This choice is made considering that the TSVD is the exact solution to the low-rank approximation problem. In other word give a random variable realization the TSVD provide the exact solution to the standard low-rank approximation problem. In a sense, the standard formulation of the standard low-rank approximation problem does not take into consideration the statistical distribution of the experimental data and considers a linear algebra perspective where the input matrix is not a random variable and the similarity between vector space elements is measured in term of Frobenoius norm without taking into account any measure of the statistical distribution of the process that generates the experimental data. In principle, if uncertainty on the elements of the matrix $\mathbf{P_1}$ is known then its effect on the residual noise could be calculated using the algebra of random variable. This calculation is more complex than the one reported in this paper.

 *5.1 Montecarlo Simulation*
Figure 2 shows that the low-rank approximation filter significantly reduces the MSE between the filtered data and the true signal, compared to the noisy signal. This reduction depends on the measurement data SNR and the number of singular values used for defining the filter. If the number of singular values used is equal to or higher than the rank of the pure signal matrix, 4 or more in this case, then the reduction in the MSE is monotonic and decreases with the number of singular values used for the filter. In this subgroup, a strong filter with few components produces a lower MSE than a weak filter with more components, because there are enough components to describe the entire signal, the pure signal has only 4 components, and the additional components end up representing noise. Therefore, in this subgroup, the four components filter is the best performing in terms of MSE.
If the number of singular values considered is equal to or lower than the rank of the pure signal matrix, 4 or less in this instance, the optimal number of components, in terms of MSE, depends on the SNR of the measurements. For each number of singular values used to build the filter an SNR interval exists in which the filter has the lowest MSE among the other low-rank approximation filters.

The data reported in figure 3 describes a general improvement, over the least square estimator (MLE) for unfiltered data, in the estimation of the effective conversion constant of lactate for both the standard least square estimator (LS-SVD, left) and the new maximum likelihood estimator (MLE-SVD, right) over the filtered data. The highest improvements are observed once the SNR level is higher than 16. For fixed SNR values, the intensity of the improvement is related to the number of singular vectors used to define the low-rank filter, the performance maxima are achieved with 3 and 4 singular values for LS-SVD (it depends on the SNR level), and 3 in MLE-SVD. For a higher number of singular values, the estimator precision tends to converge on the MLE precision proportionally to the number of singular values used for defining the low-rank filter. This convergence is smoother for the MLE-SVD estimator compared to the LS-SVD estimator, indeed there are multiple crossings in the RMSE LS-SVD curves. The behavior of the curves for the 1 and 2 singular values filter differs significantly between the LS-SVD and the MLE-SVD. The RMSE curve for the 1 singular value filter of the MLE-SVD reaches a performance plateau at an SNR level of about 8, while the 2 singular values filter curve reaches a plateau at SNR = 1024. All in all, the 2 singular values filter outperform the 1 singular value filter for the MLE-SVD.

On the contrary, for the LS-SVD the mono singular value filter outperforms the two singular values, and it also reaches a lower RMSE plateau compared to the MLE-SVD estimator on the same data. This estimator's behavior is explainable since the underline system has 4 independent components (each molecule has a different time evolution) and therefore once we project onto a one or two-dimensional pseudo-signal subspace using the low-rank filters we are losing information on how the molecule concentration changes. Despite the loss of information, the use of the low-rank filter still increases the estimation performance for specific SNR values. This apparent contradiction is due to two phenomena. First, lactate is the second most abundant metabolite in the experiment, and therefore its contribution to the dataset variance is second only to pyruvate among the metabolites. The low-rank approximation filter of rank $k$ transforms the experimental dataset into the best low-rank approximation of rank $k$ in the Frobenius sense. This also means that the filtered signal is also the low-rank matrix which maximizes the explained variance of the experimental dataset. Therefore, we can expect that the first two singular vectors contain mostly information on the pyruvate and lactate signals. So, while we are losing information on the time variation of metabolite concentration, we are mainly losing information on bicarbonate and alanine. Second, the singular vectors corresponding to the first singular values are less affected by experimental noise corruption, therefore in low SNR measurement, just a couple of singular vectors resemble the singular vector of the pure signal.

The discrepancy in the performance between LS-SVD and MLE-SVD is mainly related to the different interactions with the spectral-metabolic model; in the MLE-SVD case, the model is projected onto the same subspace of the filtered dataset, after the projection the model and the dataset lose the same information contained in the discarded singular value. On the contrary, in the LS-SVD case, the model does not get projected onto the pseudo signal subspace and it contains components that are not present in the filtered dataset. Therefore, in this case, there is an intrinsic difference between the fitted model and the pure signal present in the experimental data.

While the use of a low-rank filter causes a reduction of the RMSE of the estimated pyruvate to lactate apparent reaction constant for our dataset for several SNR levels and different numbers of singular values used to build the filter, it also causes an increase of the CRLB for the same parameter. Therefore, there is a trade-off between the maximum accuracy attenable from the estimator, the CRLB, and the effective estimation performance, the RMSE. Indeed, a maximum likelihood estimator is guaranteed to attain the CRLB only in asymptotic behaviour. Also, the estimator is a maximum likelihood estimator only if the model perfectly matches the experimental pure signal and the experimental noise matches the experimenter's assumption. For example, the presence of a line shape distortion or a baseline makes invalid the assumption for defining the maximum likelihood estimator.

The reduction in the RMSE is not equal for all the parameters, for example in the case of the effective $T_1$ relaxation rate of bicarbonate the use of a low-rank approximation filter reduces the quality of the estimation for both estimators unless the starting SNR is higher than 1000. Such high SNR is commonly obtained in successful hyperpolarization experiments, therefore the use of a low-rank approximation filter in datasets obtained using hyperpolarized tracers should be risk-free. Nevertheless, this example should raise awareness around the limitation of the combined use of sophisticated data analysis and low-rank relaxation filters. Indeed, the only theoretical result supporting the wide use of a low-rank filter regard is its ability in reducing the MSE between the denoised data and the noiseless signal. Montecarlo simulation, such as that employed in this article, should be used to quantify the effect of the low-rank approximation filter on the data analysis pipeline.

These results have also practical implications for the design of experiments involving hyperpolarized tracers. Indeed, we have shown that low-rank filters can produce a reduction in RMSE that is greater than that which can be achieved by doubling the SNR and fitting the unfiltered data. Doubling the SNR level of a d-DNP experiment is not a trivial task; the standard solution to the sensitivity problem in an NMR experiment is increasing the number of acquisitions, this is not possible in a one-shot experiment like this. Therefore, the only solutions are: a) increase the hardware performance and reduce the electrical noise, b) increase the tracer dose, c) improve the solid-state hyperpolarization level, or d) reduce the polarization loss before the injection. None of these methods is as simple and easy as the use of low-rank approximation filters.

Therefore, the use of low-rank filters could be useful also for hyperpolarization experiments where the previous SNR reduction strategies are difficult to implement such as methods based on para-hydrogen[140] due to the lower polarization levels achieved compared to d-DNP or with long-time stored hyperpolarized sample[141].

**6 Conclusions**

In this paper, we presented an alternative interpretation of the low-rank approximation filter as a linear filter with a rank deficit matrix representation. The mean and variance of the filtered data are derived using this interpretation and the properties of multilinear functions. We also showed that data with multivariate normal distribution, such as a deterministic signal corrupted by additive Gaussian noise, remain normally distributed after the use of such a filter.

These results can be used to improve the analysis of the filtered data, for example, by defining a maximum likelihood estimator for the filtered data.

Three different estimators were compared, i.e. the previously mentioned maximum likelihood estimator on the filtered data (MLE-SVD), the least square estimator on the filtered data (LS-SVD), and the maximum likelihood estimator on the unfiltered data (MLE). In particular, the effectiveness of the low-rank filter in improving the NMR data analysis was evaluated in a Montecarlo simulation of hyperpolarized d-DNP experiments, especially keeping in mind that such improvement should primarily affect the accuracy and precision of the information derived from the data.

The simulation converted multiple SNR levels over three orders of magnitude and different approximation ranks. We focused our analysis on the reaction rate of the metabolic conversion of pyruvate into lactate and bicarbonate due to its importance for studying cell metabolism. It was found that the use of a low-rank approximation filter significantly improves the RMSE compared to the unfiltered case for all SNRs higher than 16. Also, the MLE-SVD performance is more consistent with respect to the LS-SVD estimator along all the SNR levels and different reconstruction ranks. Unfortunately, the use of a low-rank approximation filter increases the CRLB for the MLE estimator. This increment is higher for low SNR levels and decreases to almost zero for SNR higher than 128.

Therefore, there is a trade-off between the decrease in the RMSE in the empirical parameter estimation and the reduction in potential precision measured as an increase of the CRLB.
For the effective spin-lattice relaxation time of bicarbonate, the use of a low-rank approximation filter increases the RMSE for almost all SNRs lower than 1024. This effect is stronger in the LS-SVD estimator compared to the MLE-SVD one.
Overall, we suggest testing the effect of the low-rank approximation filter on a Montecarlo replica of the experimental data to evaluate the benefits of using such a filter along with the subsequent analysis.
To conclude, low-rank approximation filters greatly enhance the separation between the signal and the noise, measured as the SNR, and the fidelity between the filtered signal and the pure signal, measured as the MSE. Nevertheless, this family of filters has a profound impact on the filtered signal statistics changing both the mean and the covariance matrix of the filtered data. This paper describes a method to quantify these changes. This additional knowledge on the filtered dataset must be used to fully exploit the noise reduction properties of low-rank approximation filters.

**Appendix A:**

**Proposition 1:** Let $C_1 = \sigma^2 \, TIT^*$ were $T = (P_1 \otimes I)$ and $I$ is the identity matrix, all matrices belong to $\mathbb{C}^{mn \times mn}$. Then $C_1 = \sigma^2 (P_1 P_1^*) \otimes I$.

Proof:
$$C_1 = \sigma^2 (P_1 \otimes I)(P_1 \otimes I)^* =$$
$$= \sigma^2 (P_1 \otimes I)(P_1^* \otimes I^*) =$$
$$= \sigma^2 (P_1 P_1^*) \otimes (II^*) = \sigma^2 (P_1 P_1^*) \otimes I$$

**Proposition 2:** The scalar $-1/\sigma^2 (a^* T^*) C_1^+ (Ta)$ is equal to $-1/\sigma^2 \|DA\|_F$, where $a$ belongs to $\mathbb{C}^{mn}$, and $T$ and $C_1$ belong to $\mathbb{C}^{mn \times mn}$ assuming that: $T^*T = C_1$, $T^+T = (P_1^+ P_1) \otimes I$, $T^+Ta = P_1^+ P_1 A$, and $T = (P_1 \otimes I)$.

Proof:
$$-1/\sigma^2 (a^* T^*) C_1^+ (Ta) =$$
$$= -1/\sigma^2 (a^* T^*)(TT^*)^+ (Ta) =$$
$$= -1/\sigma^2 (a^* T^*) T^{*+} T^+ (Ta) =$$
$$= -1/\sigma^2 (T^+ Ta)^* (T^+ Ta) =$$
$$= -1/\sigma^2 (P_1^+ P_1 A)^* (P_1^+ P_1 A) =$$
$$-1/\sigma^2 (DA)^* (DA)$$

Where $D$ is equal to $P_1^+P_1$. This last therm is equal to minus the Frobenius norm of matrix $DA$ divided by sigma.

**Proposition 3:** If the matrix P1 is an orthogonal projection than $P_1^+ P_1 = P_1$

Proof: If P1 is an orthogonal projection then $P_1 = BB^+$ where $B$ is an opportune matrix, then using Penrose inverse properties $P_1^+ P_1 = (BB^+)^+ (BB^+) = BB^+ BB^+ = BB^+ = P_1$.

**Supplementary Material:**
The supplementary material is divided into three parts: the first provides a detailed description of the generalization of the paper results to HOSVD or Hankel matrix-based filter, the second one contains the results of the Montecarlo simulation for all the model's parameters, the last part provides a qualitative description of the effects of low-rank approximation filter on the synthetic datasets.

**Data availability:**
All the results from the Montecarlo simulation are available upon request from the authors both as .txt and .npy files.
The code to produce the plots and the Montecarlo simulation is available upon reasonable request from the authors.

**Conflict of Interest Statement:** The authors have no conflicts of interest to declare
**Declaration of generative AI in scientific writing:** The authors did not use any AI tool in the writing of this paper.

# Supplementary material

# Contents

# List of Figures

# 1 Introduction

This documents is divided in three sections:the first one describe how to use the results from the main paper to study the residual noise of low rank approximation filter for non matrix data, the second one contains the results of the Montecarlo simulation for all the model parameters, the third one contains a qualitative description of the simulated datasets and its analysis with low-rank approximation filter.

All the images, and the caption reported in this document where previously published by Roberto Francischello in the Ph.D. thesis *"Development of new experimental and data processing methods at critical signal-to-noise conditions in nuclear magnetic resonance"*, and here reproduced with his permission.

# 2 Residual noise characterization for non matrix datasets

## 2.1 Low-rank approximation filter based on Higher Order Singular Values Decomposition

In the main article we provided a brief explanation of how HOSVD could be use to build low-rank approximation filter for multi-dimensional datasets. This is the most common approach to the low-rank filter of tensor data with additive noise in MRI. In the following, we try to make the whole procedure more accessible by using an informal description of the HOSVD procedure.

The algorithm to calculate the HOSVD is:

- Unfold the tensor along the $k$ axis producing a matrix $M_k$

- Calculate the SVD of the matrix $\mathbf{M}_k$ and store the left singular vector matrix $\mathbf{U}_k$ e the singular values $\mathbf{S}_k$

- Repeat from point 1 until all the axis of the tensor are used

- Construct the Tensor core $\mathcal{C} = (\mathbf{U}_1, \ldots, \mathbf{U}_n) \, dot \mathcal{A}$, where $dot$ is the multi-linear multiplication

A low-rank approximation of the tensor $\mathcal{A}$ is built by zeroing some entry of the core tensor, and the singular matrix Sk can be used to decide which term to withhold. Unfortunately, there isn't an Eckart-Young theorem generalization for the low-rank approximation of tensor and therefore the approximation is not optimal.

The standard procedure for the calculation of the THOSVD is equivalent to:

- Unfold the tensor into a 1-D vector

- Fold the 1-D vector $L_k$ into one of the $\mathbf{M}_k$ unfolding matrices

- Apply the linear projection $\mathbf{U}_k \mathbf{U}_k^\dagger$ to the $\mathbf{M}_k$ matrix

- Vectorize the matrix $\mathbf{M}_k$

- Apply a permutation, $\pi_k$, to the 1-D vector such as $\pi_k \mathbf{L}_k = vec(\mathbf{M}_{k+1})$

- Repeat from point 2 until all the axis of the tensor are used

Therefore, the final version of the data is obtained by a consecutive use of linear transform (linear projection and permutation), vectorization, and matricization.

If $\pi_1, \pi_2, \ldots \pi_{n-1}$ are the permutation matrix needed to obtain all the possible matrix configurations of the tensor $\mathcal{A}$ and $\mathbf{Q}_1$ to $\mathbf{Q}_n$ are all the projection matrix needed to obtain truncated HOSVD approximation of the tensor A, and the transform matrix $\mathbf{T}$ is equal to the product $\mathbf{T} = \mathbf{Q}_n \pi_{n-1} \mathbf{Q}_{n-1} \pi_{n-2} \ldots \pi_1 \mathbf{Q}_1$, than $\mathbf{T} vec(\mathcal{A}) = vec(\hat{\mathcal{A}})$ where $\hat{\mathcal{A}}$ is a low-rank approximation of the tensor $\mathcal{A}$ obtained with the THOSVD. Therefore, we can apply the main results of the paper and calculate the statistical distribution of the filtered data using the linear transformation which is equal to the THOSVD procedure. This procedure can be applied both to global low-rank filters and patch-based low-rank filters or any combination of the two. Patch-based local filters require extreme carefulness in the aggregation phase, the final value for the voxel in position $(x1, y1, z1)$ is the weighted mean of all the values of voxel $(x1, y1, z1)$ in the different patch. Therefore, the mean and standard deviation of the denoised signal is just an accounting exercise using the linearity of the mean and the bi-linearity of the variance.

## 2.2 Low-rank approximation filter based on Hankel matrix

To treat mono-dimensional data using a low-rank approximation filter the original measured signal should be rearranged into a matrix form to exploit a latent low-rank structure. The most common procedure is the form a Hankel matrix, this particular matrix has a low-rank structure if the original signal is autoregressive. The Cadzow filter and the Singular Spectrum Analysis are the most used Hankel-based low-rank approximation filter. A full description of both methods goes far beyond the scope of this appendix, therefore we refer interested readers to the specialistic literature.

Both methods start with the same algorithm:

- Forming the Hankel matrix $\mathbf{H}$ from the original data

- Calculate the SVD of $\mathbf{H}$

- Select the reconstruction rank and build the $\hat{\mathbf{H}}$ low-rank matrix

- Reconstruct the mono-dimensional signal by the anti-diagonal average of the matrix $\hat{\mathbf{H}}$

In the Cadzow filter, the procedure is repeated until the Hankel matrix is already of the right rank.

The main result from the article covers the characterization of the residual noise from step 1 to step 3. Therefore, after the third step the signal, after its vectorization, has the mean $\mu$ and the covariance matrix $\mathbf{C}$ calculated according to the main result of the paper. Due to the average over the antidiagonal, each element $x_t$ of the final signal is equal to

$$x_t = \frac{1}{k_t} \sum_{ij} h_{ij}, \quad i+j = k_t$$

Where $h_{ij}$ is an element of the low-rank approximation matrix, $k_t$ is the number of the elements of the $t^{th}$ anti-diagonal.

The mean of the $x_t$, due to the linearity of the operator of the expected value is:

$$\bar{x}_t = \frac{1}{k_t} \sum_{ij} \bar{H}_{ij}, \quad i+j = k_t$$

$$\bar{x}_t = \frac{1}{k_t} \sum_{ij} \mu_{ij}, \quad i+j = k_t$$

The new covariance matrix $\hat{\mathbf{C}}$, of the one dimensional signal, due to the bi-linearity of the operator is:

$$\hat{C}_{t_1 t_2} = cov\left[x_{t_1}, x_{t_2}\right]$$

$$\hat{C}_{t_1 t_2} = cov\left[\frac{1}{k_{t1}} \sum_{i+j=k_{t_1}} H_{ij}, \frac{1}{k_{t2}} \sum_{l+m=k_{t_2}} H_{lm}\right]$$

$$\hat{C}_{t_1 t_2} = \frac{1}{k_{t1}} \frac{1}{k_{t2}} \sum_{i+j=k_{t-1}} \sum_{l+m=k_{t_2}} cov\left[H_{ij}, H_{lm}\right]$$

The covariance of elements $\mathbf{H}_{ij}$ and $\mathbf{H}_{lm}$ are just the corresponding entry in the covariance matrix (after the low-rank approximation) $\mathbf{C}$ depending on the vectorization operation used. If $\mathbf{H}$ is an $N \times M$ matrix, with row first vectorization the elements $H_{ij}$ correspond to the element $h_{(i-1)M+j}$ after the vectorization. Therefore, the $cov[\mathbf{H}_{ij}, \mathbf{H}_{lm}]$ is equal to the elements $((i-1)M+j, (l-1)M+m)$ of the covariance matrix $\mathbf{C}$ of the low-rank approximation matrix obtained at the third step of the algorithm.

If the starting noise is Gaussian additive noise, the residual noise after the use of the filter is still Gaussian, indeed all the operations applied from steps 1 to 4 are linear. Nevertheless, the final noise is highly correlated even if the original noise is white. In principle, the knowledge of the covariance matrix after the low-rank approximation could be used to optimize the covariance matrix of the data after the anti-diagonal average procedure. If the simple average along the anti-diagonal is substituted with a weighted average, the final correlation matrix became a function of the average coefficients. Therefore, it is possible to minimize the SNR of the final signal or minimize the norm of the covariance matrix.

Some Hankel matrix-based low-rank approximation methods, such as LORA, do not have anti-diagonal average phases, they simply use the first row and the last columns of the matrix to construct the final denoised signal. Therefore, the statistical distribution of the residual noise is directly calculated using the main results from the article.

## 3  Singular values shrinkage filters

Singular values shrinking, reducing the values of singular values based on on heuristic or theoretical results, aims to reduce the distance between the true signal matrix and the low-rank approximation of the noisy measurements matrix. Among those methods the one based on random matrix theory are the most theoretically robust, they start with the SVD of the measurements matrix and define a function f(x) that reduces the intensity of the singular values according to random matrix theory to minimize the distance between the approximation matrix and the noiseless signal matrix.

The results of the singular values shrinking procedure are equivalent to the linear transform of the measurement matrix with the linear transform $\mathbf{\Psi} = \mathbf{U}\mathbf{H}\mathbf{U}^H$, where $\mathbf{H}$ is a diagonal matrix whose diagonal elements are the shrinking coefficients. Then the transformed signal is equal to :

$$\mathbf{\Psi M} = \mathbf{\Psi U \Sigma V} \tag{1}$$
$$= \mathbf{U H U}^H \mathbf{U \Sigma V} \tag{2}$$
$$= \mathbf{U H \Sigma V} \tag{3}$$
$$= \mathbf{U \Sigma}_\eta \mathbf{V} \tag{4}$$
$$= \mathbf{M}_\eta \tag{5}$$

Unlike the standard low-rank approximation problem this transform is not a projection matrix. Nevertheless, almost all the results reported in the article's main body still hold. The mean and the covariance matrix of the filtered data are still obtained from the original mean and covariance matrix as:

$$\bar{\boldsymbol{\mu}} = \mathbf{\Psi}\boldsymbol{\mu} \tag{6}$$
$$\bar{\mathbf{C}} = \mathbf{\Psi C \Psi} \tag{7}$$

Also, Gaussian-distributed data are still Gaussian-distributed after the application of the filter. Finally, the use of singular values shrinking does not change the MLE estimator compared with the standard low-rank approximation filter with the same rank. Indeed, the MLE loss function depends on the products of the inverse of the transform matrix, with the transform matrix itself $\mathbf{T}^+\mathbf{T}$.

When the filter is the standard low-rank approximation $\mathbf{T} = \mathbf{P}$, where $\mathbf{P}$ is the projection matrix when the filter is the singular values shrinking one $\mathbf{T} = \mathbf{\Psi}$.

$$\mathbf{\Psi}^+\mathbf{\Psi} = (\mathbf{UHU}^*)^+ \mathbf{UHU}^* \tag{8}$$
$$= \mathbf{U}^{*+}\mathbf{H}^+\mathbf{U}^+\mathbf{UHU}^* \tag{9}$$
$$= \mathbf{U}^{*+}\mathbf{H}^+\mathbf{HU}^* \tag{10}$$
$$= \mathbf{U}^{*+}\mathbf{I}_r\mathbf{U}^* \tag{11}$$

$\mathbf{H}$ is a diagonal matrix, therefore $\mathbf{H}^+\mathbf{H}$ is a diagonal matrix, $\mathbf{I}_r$ with $r$ ones along the diagonal, corresponding to the non-zero elements of $\mathbf{H}$, and zeros otherwise.

This results is equal to that obtained for $\mathbf{P}^+\mathbf{P}$:

$$\mathbf{P}^+\mathbf{P} = (\mathbf{UI}_r\mathbf{U}^*)^+ \mathbf{UI}_r\mathbf{U}^* \tag{12}$$
$$= \mathbf{U}^{*+}\mathbf{I}_r^+\mathbf{U}^+\mathbf{UI}_r\mathbf{U}^* \tag{13}$$
$$= \mathbf{U}^{*+}\mathbf{I}_r^+\mathbf{I}_r\mathbf{U}^* \tag{14}$$
$$= \mathbf{U}^{*+}\mathbf{I}_r\mathbf{U}^* \tag{15}$$

Since the loss function is equal for both filters the solution to the MLE should be equal. Nevertheless, this result holds only for MLE with additive Gaussian noise, the use of singular values shrinking still changes the mean and the covariance matrix of the residual noise, therefore it can be a useful tool when combined with other kinds of estimators, like peak integration, that have a stronger dependency on the MSE between the noiseless signal and the filtered one.

# 4 Additional results for the properties of the maximum likelihood estimator on filtered data

### Results form the Montecarlo simulation of model fitting

This section contains the full results on the estimator performance from the Montecarlo simulation. There are 17 panels, one for each parameter.

The first row of the panels present the comparison of the Mean Square Error (instead of the normalized Root Mean Square Error) on the estimated parameter for the three estimators studied. Left: comparison between the least squares estimation on the raw data[1], bright green star, and the least squares estimation on the filtered data, solid lines. Right: comparison between the least squares estimation on the raw data, bright green star, and the maximum likelihood estimation on the filtered data, solid lines. All the axes are in logarithmic scale.

The second row of the panels present the comparison between the p-values obtained using the Wilcoxon signed rank test to test the hypothesis "The difference between the absolute estimation error for the Filtered Least Square (FLS) or the Filtered Maximum Likelihood (FML) estimator and the one for the reference estimator the Original Least Square (OLS) have zero median". We reported on the abscissa the number of singular values used to define the filter and on the ordinate the SNR of the signal. On the left the result for comparison between the FLS and the reference on the right the comparison between the FML and the reference are reported.

---

[1] wich is also the maximum likelihood estimator because the noise is white and Gaussian

Finally, the third row of the panels present the comparison of the Common Language Effect Size for the absolute error in the estimation of the parameters using the three estimators studied. We compare, pair wise, the performance of the estimation on filtered data, both in the least squares and in maximum likelihood framework, with the least squares estimation on the untreated data, as our reference. The CLES is equal to the fraction of estimation that has less absolute error than the reference. We present our data using a heat map, reporting on the abscissa the number of singular values used for defining the filter, and on the ordinate the SNR level. Bright orange means that in a higher number of pairs the estimator working on the filtered data has less absolute error compared with the reference, bright green means that the reference estimator provides a better estimation.

## Estimated Cramer Rao Lower Bound

Here we report the estimated values of the CRLB for the diagonal elements of the covariance matrix. The CRLB are reported in 4 panels, each panel reports the CRLB of the four parameters that describe that model the behavior of each molecule.

The estimation of the CRLB on the raw data, bright green star, and the CRLB estimation on the filtered data, solid lines. The number of singular values used to build the filter is color coded, from the single singular value of the yellow line to the 59 singular values of the dark purple line. The discontinuities in the line are caused by improper value for the CRLB, the CRLB is always positive while the values were negative, probably due to numerical errors. The origin of these errors is discussed in the main body of the article.

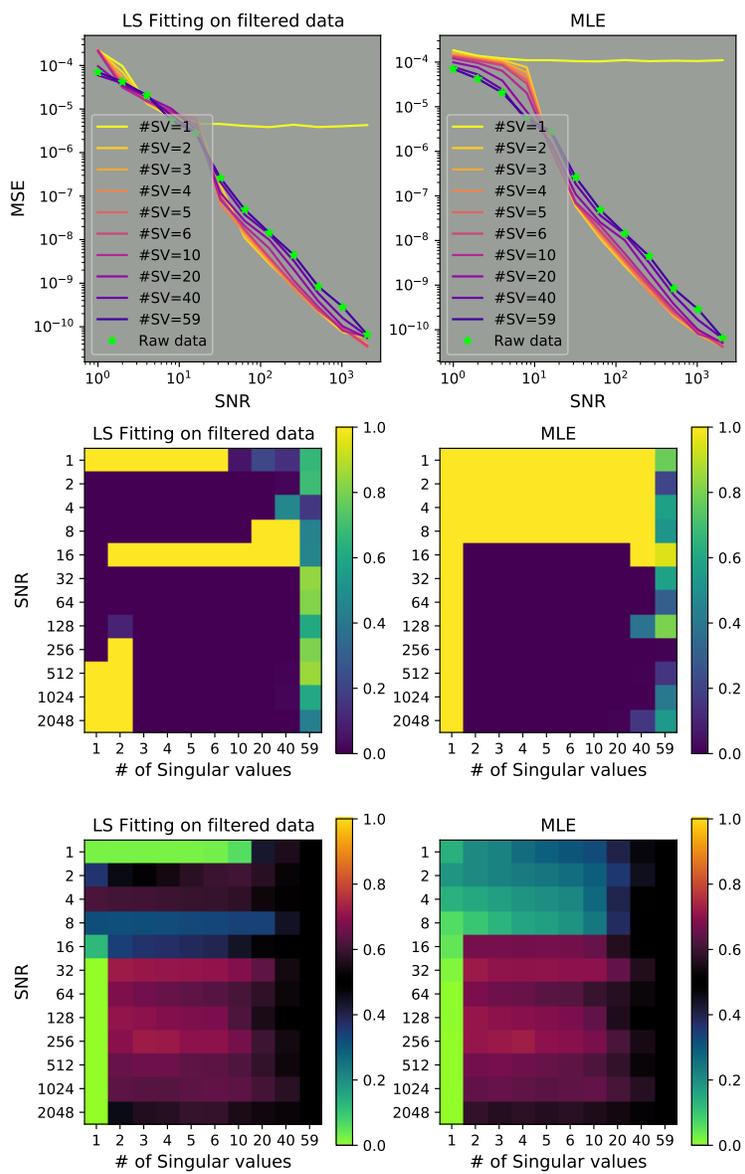

Figure 1: Pyruvate $T_{1eff}$ relaxation rate

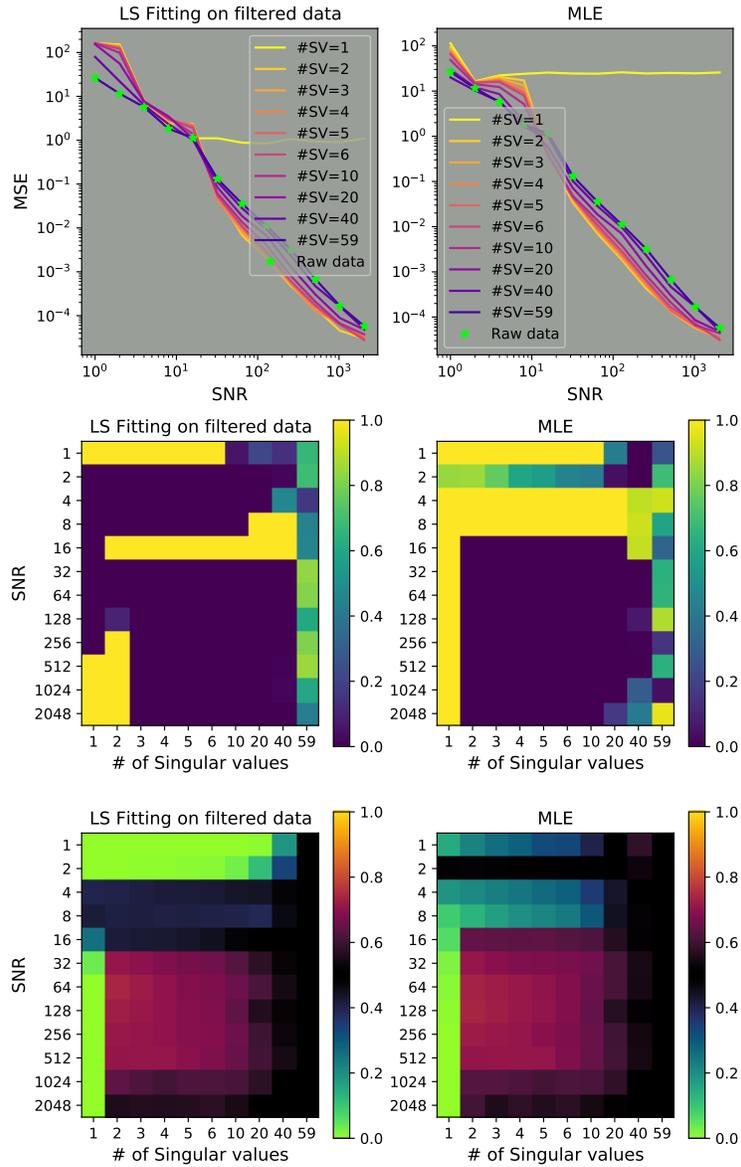

Figure 2: Pyruvate injection rate

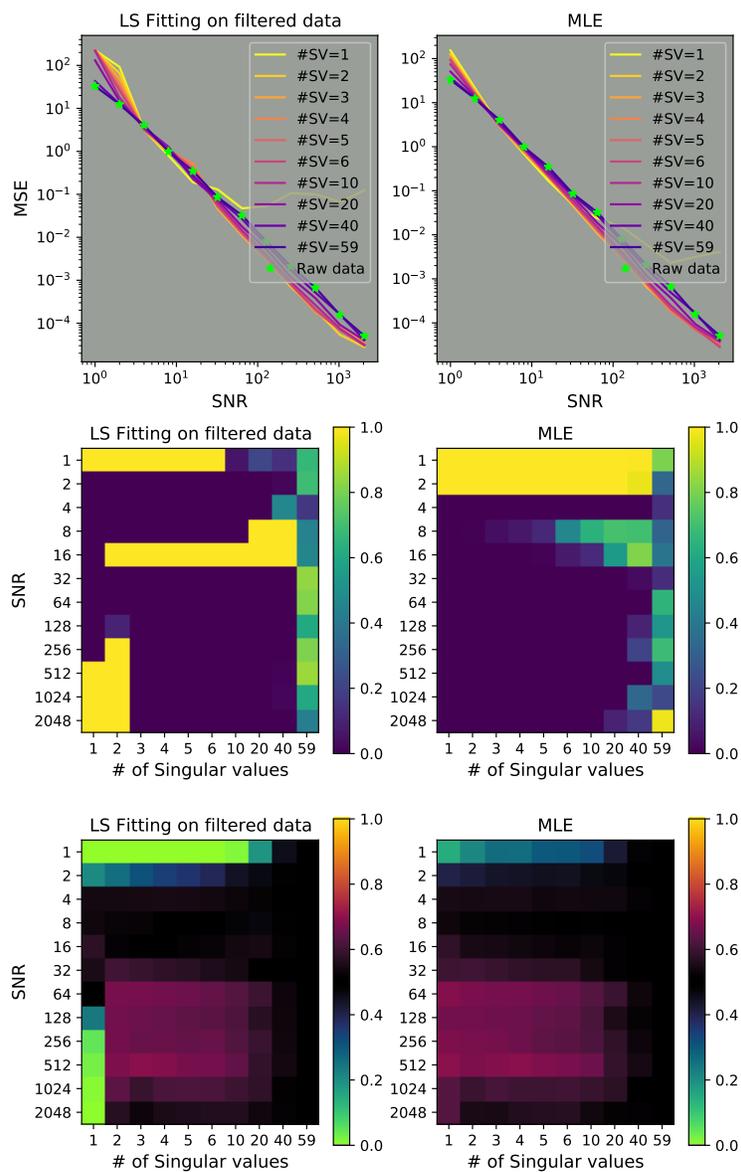

Figure 3: Pyruvate $T_2$ relaxation rate

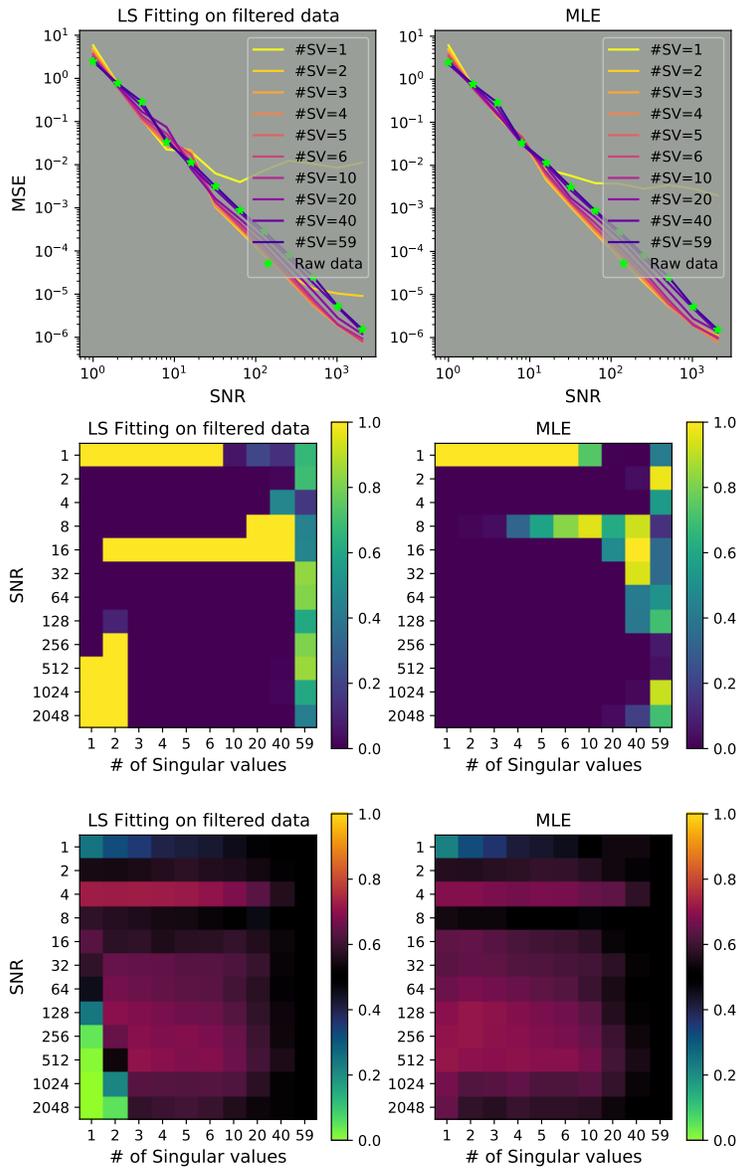

Figure 4: Pyruvate frequency

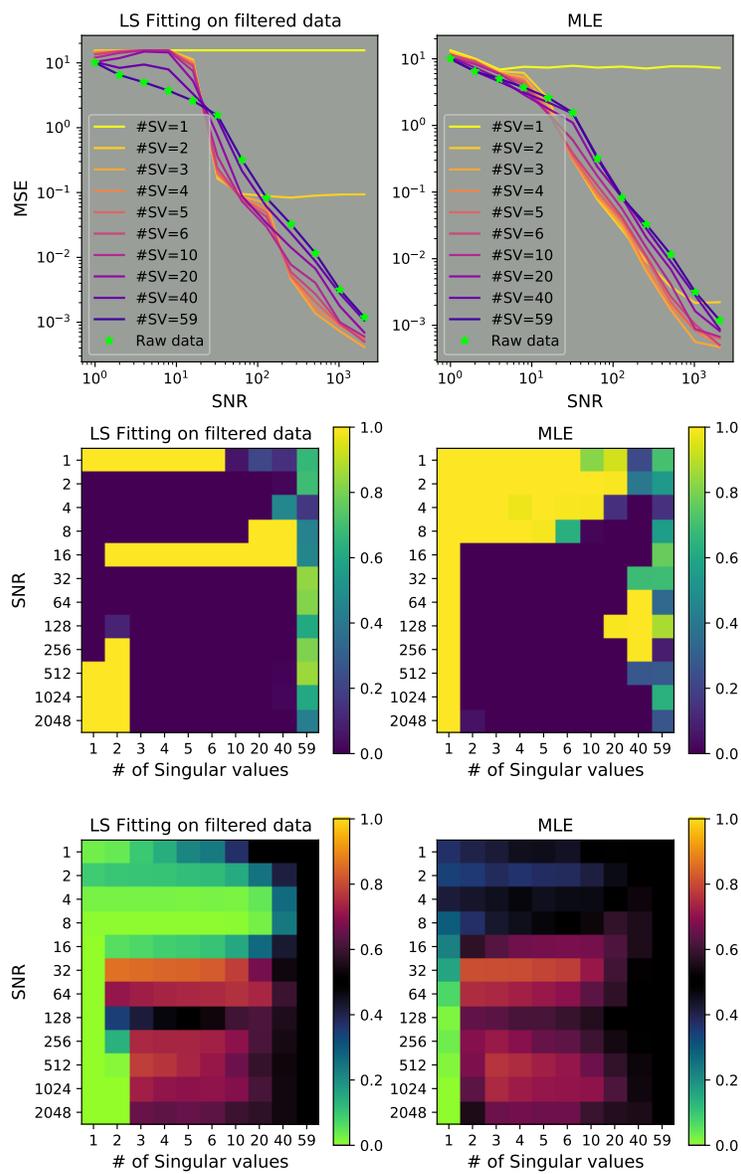

Figure 5: Lactate $T_{1eff}$ relaxation rate

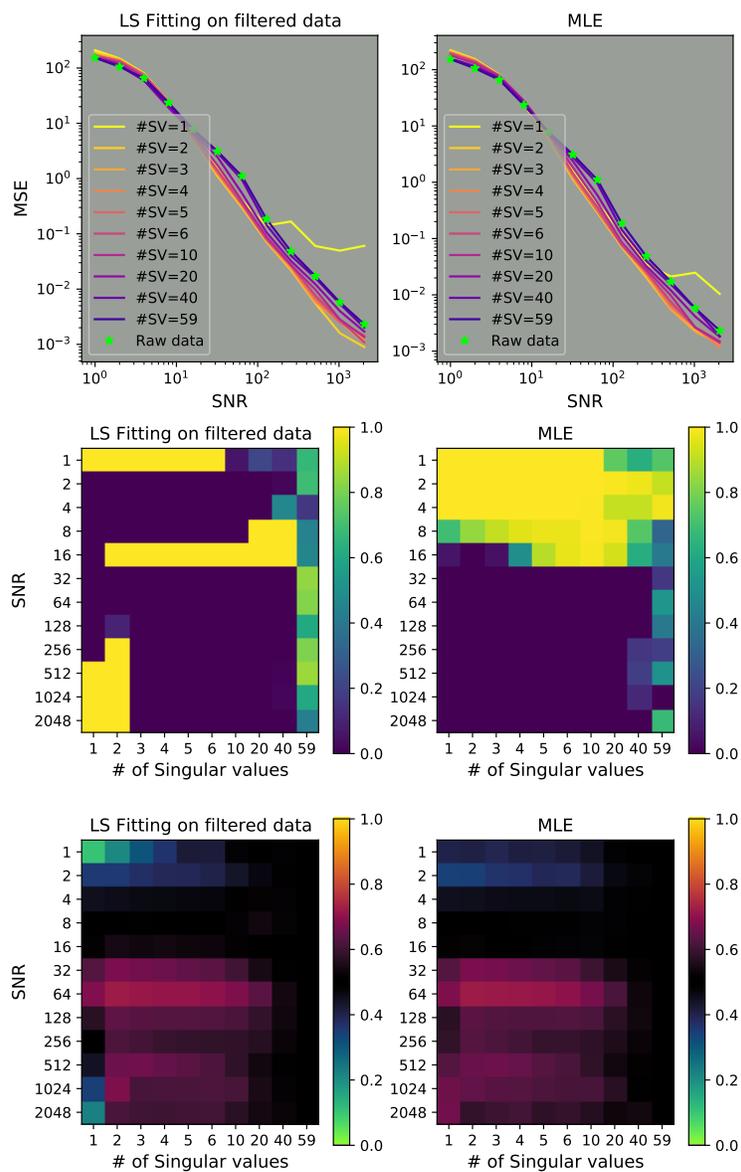

Figure 6: Lactate $T_2$ relaxation rate

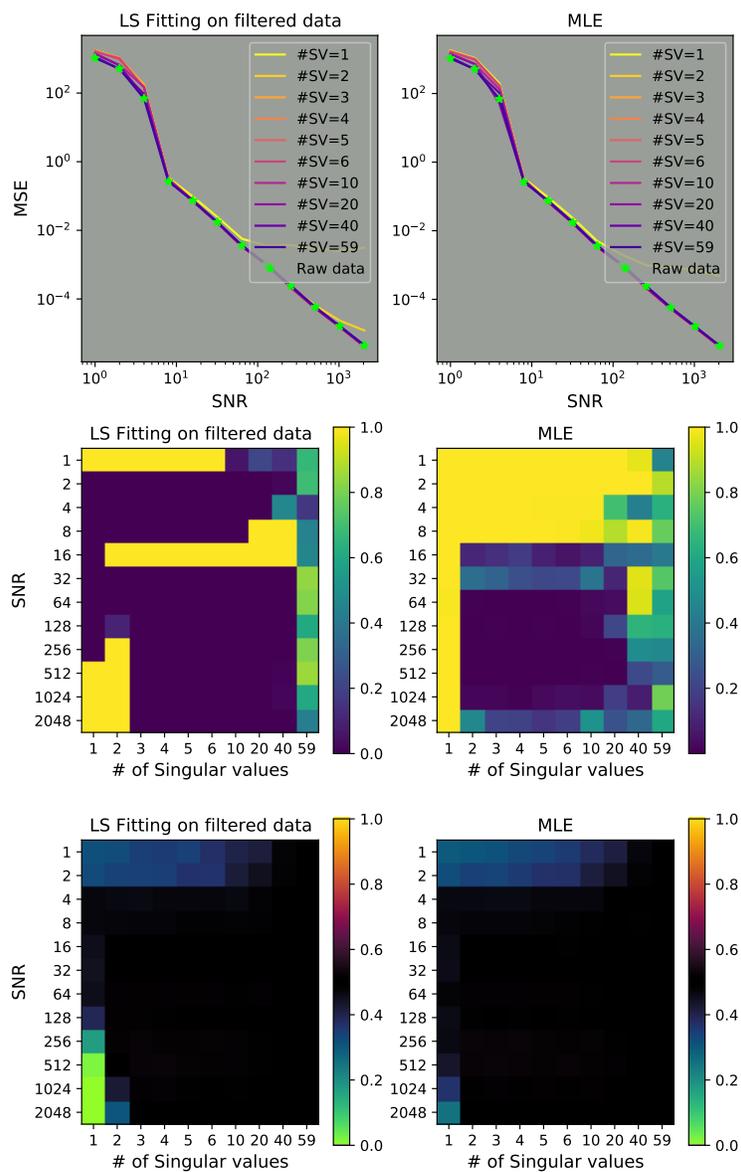

Figure 7: Lactate frequency

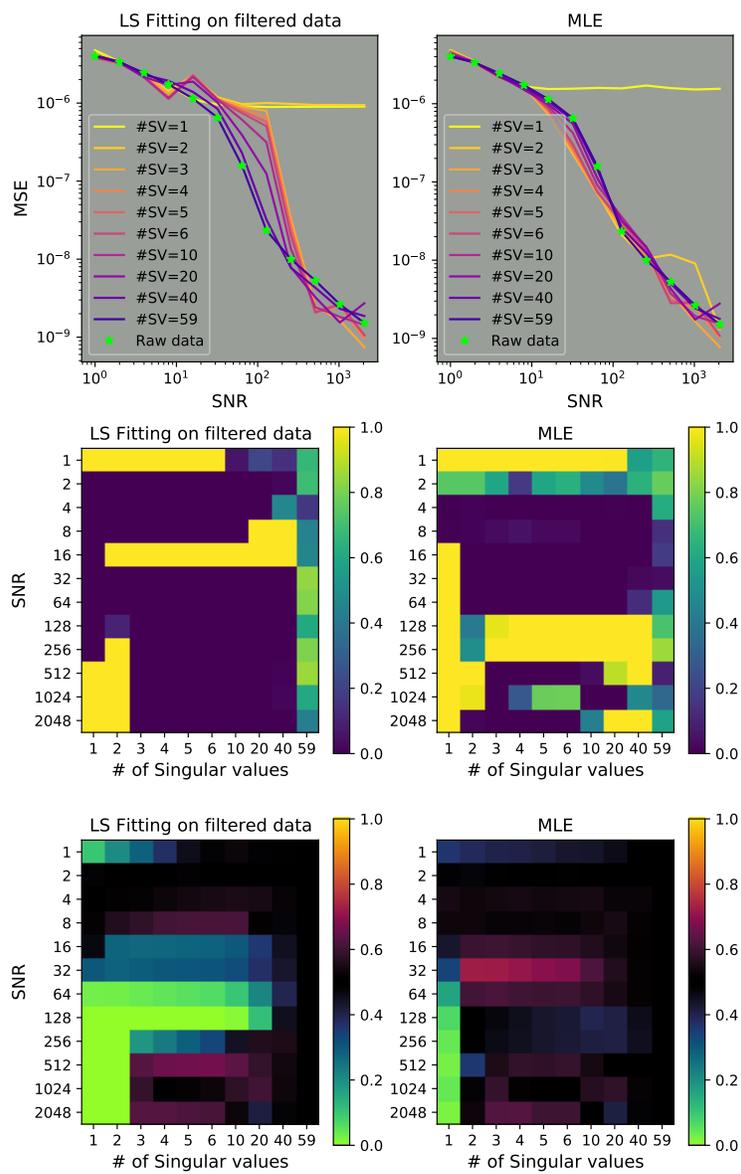

Figure 8: Bicarbonate effective reaction rate

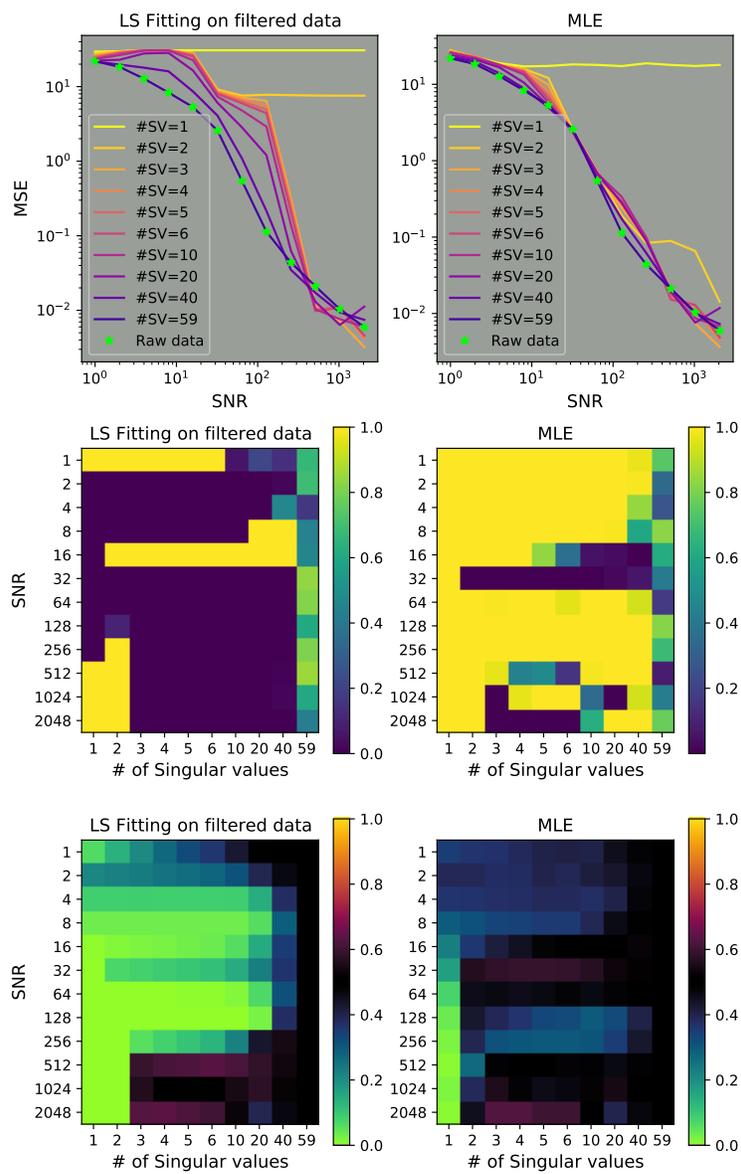

Figure 9: Bicarbonate $T_{1eff}$ relaxation rate

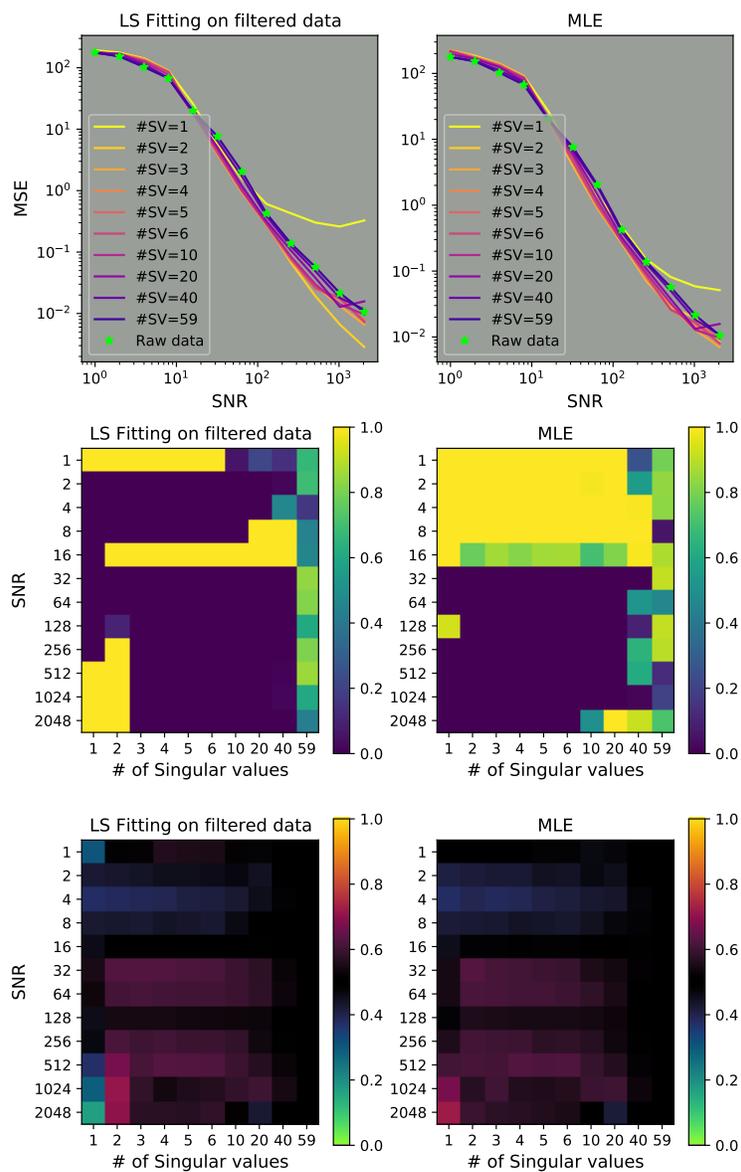

Figure 10: Bicarbonate $T_2$ relaxation rate

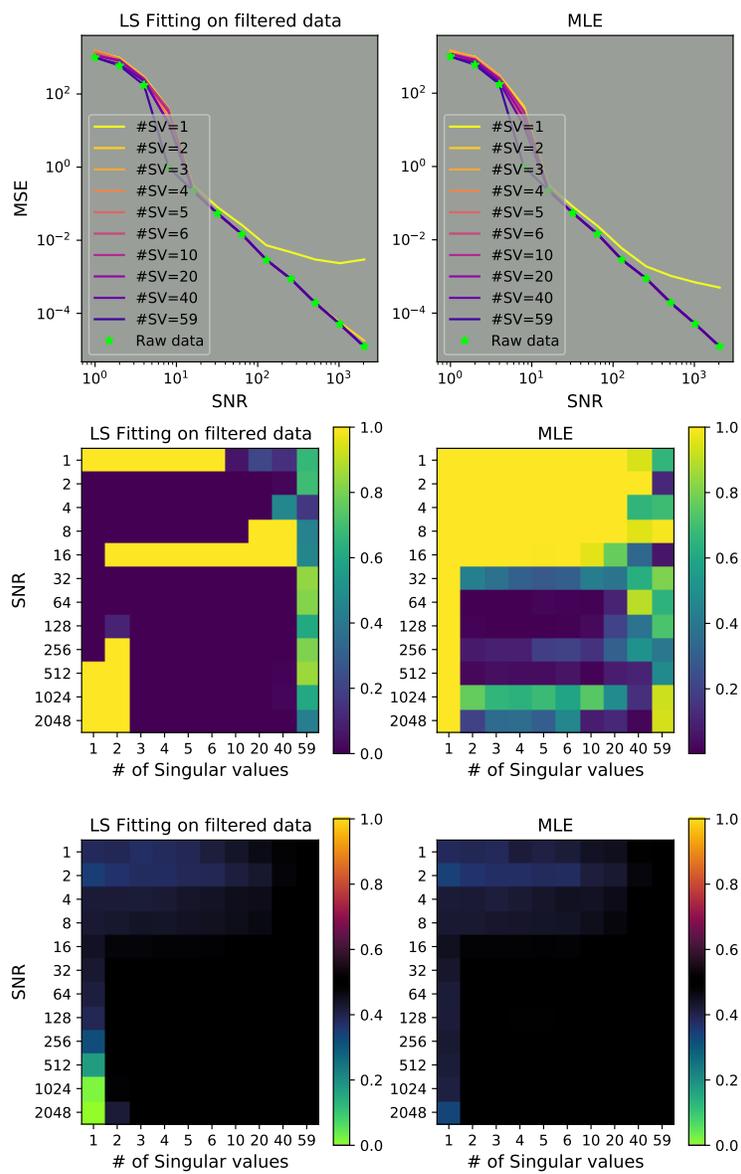

Figure 11: Bicarbonate frequency

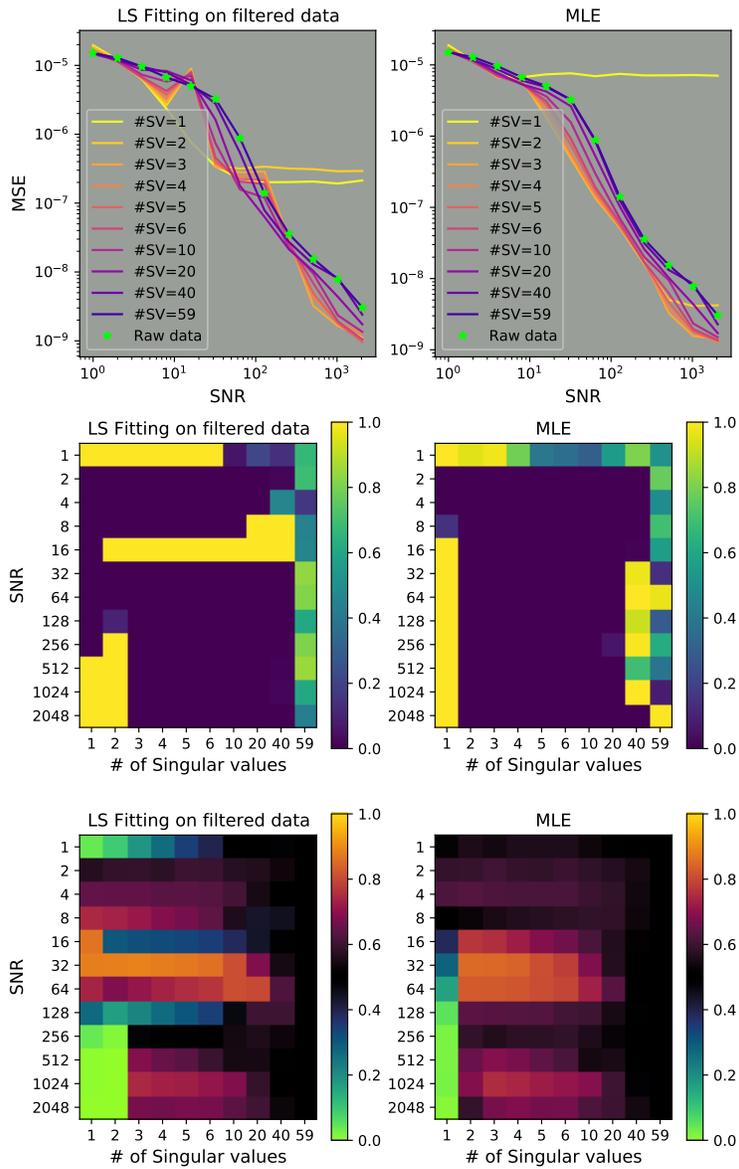

Figure 12: Alanine effective reaction rate

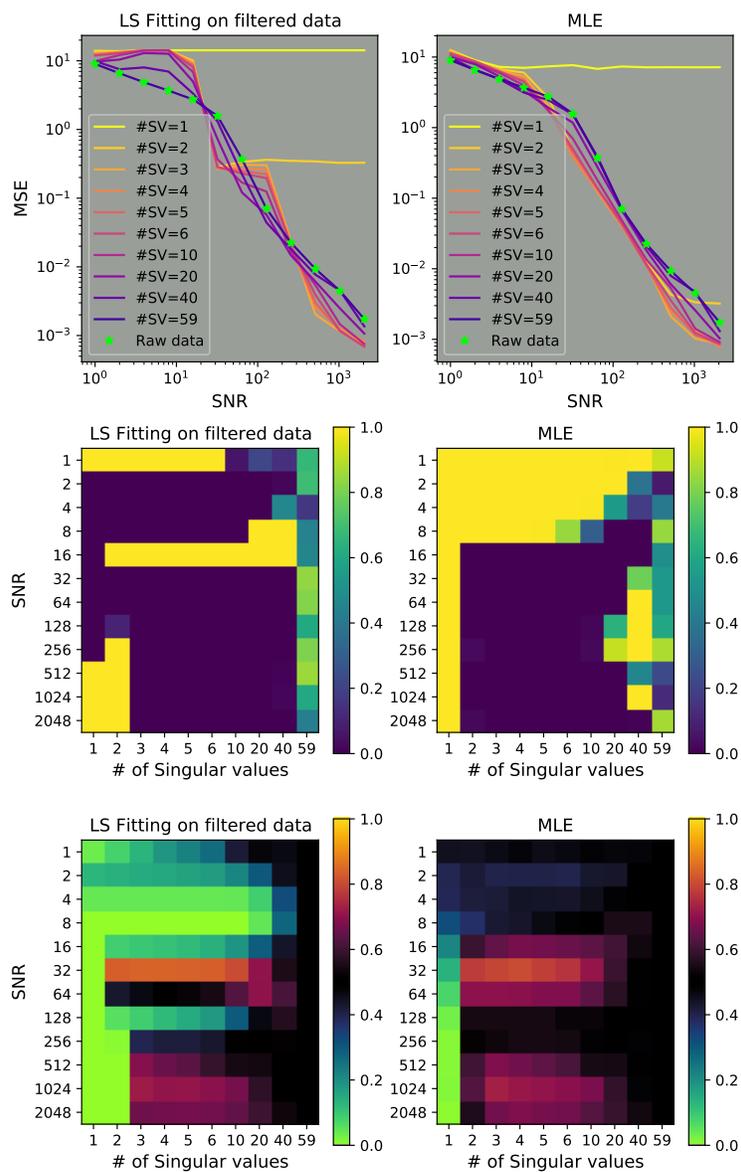

Figure 13: Alanine $T_{1eff}$ relaxation rate

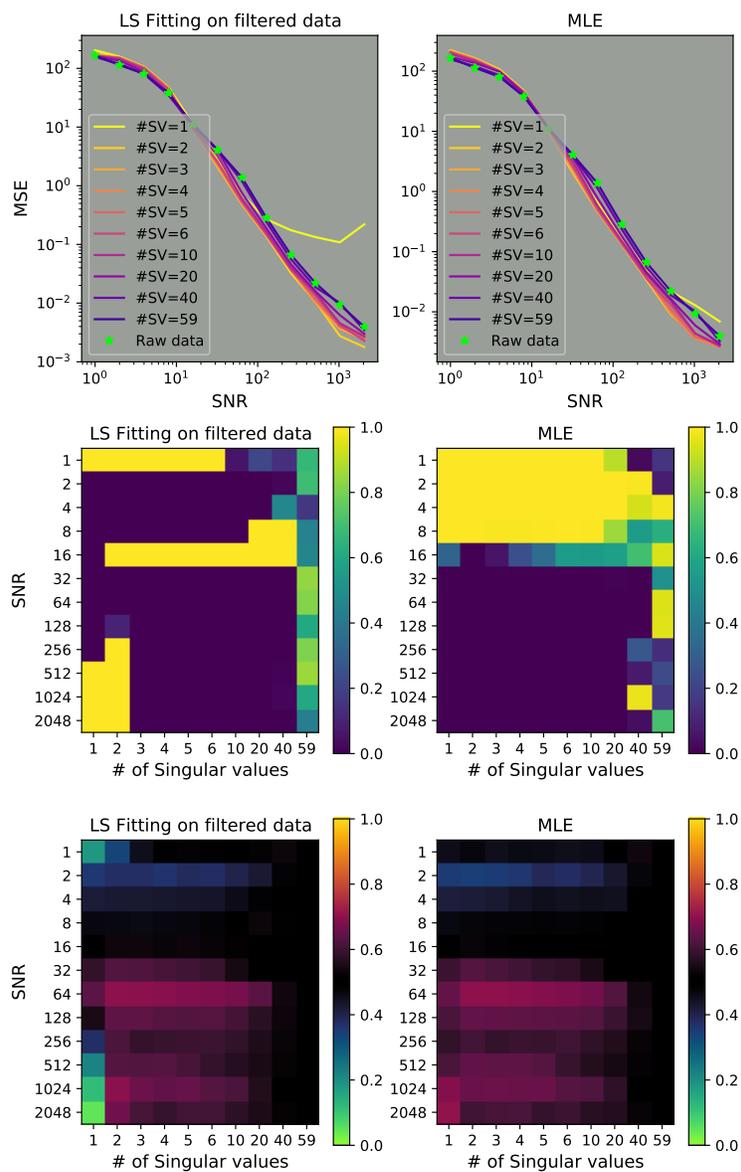

Figure 14: Alanine T$_2$ relaxation rate

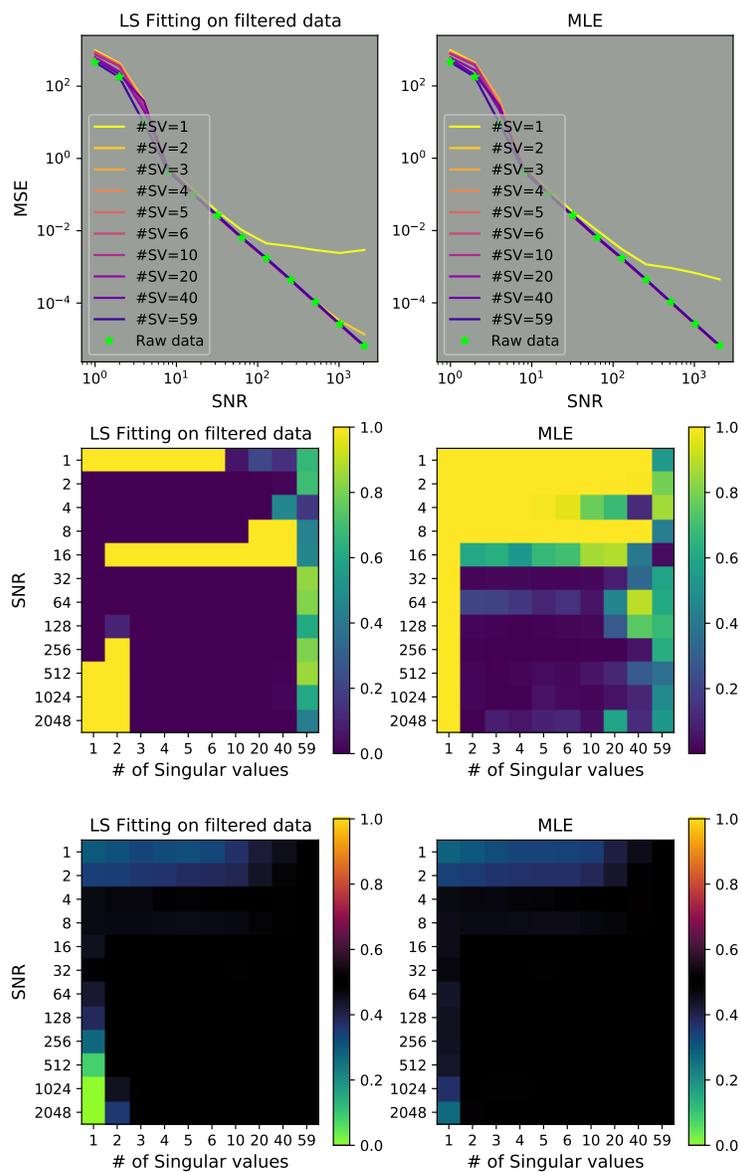

Figure 15: Alanine frequency

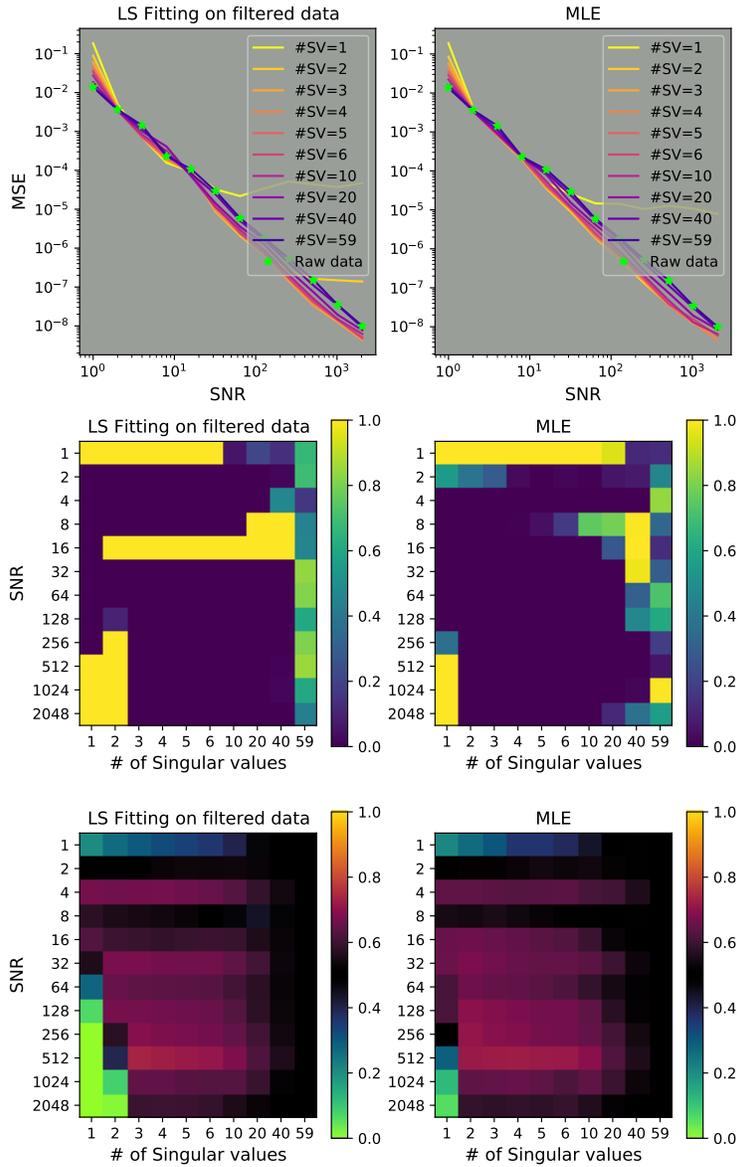

Figure 16: Zero order phase factor

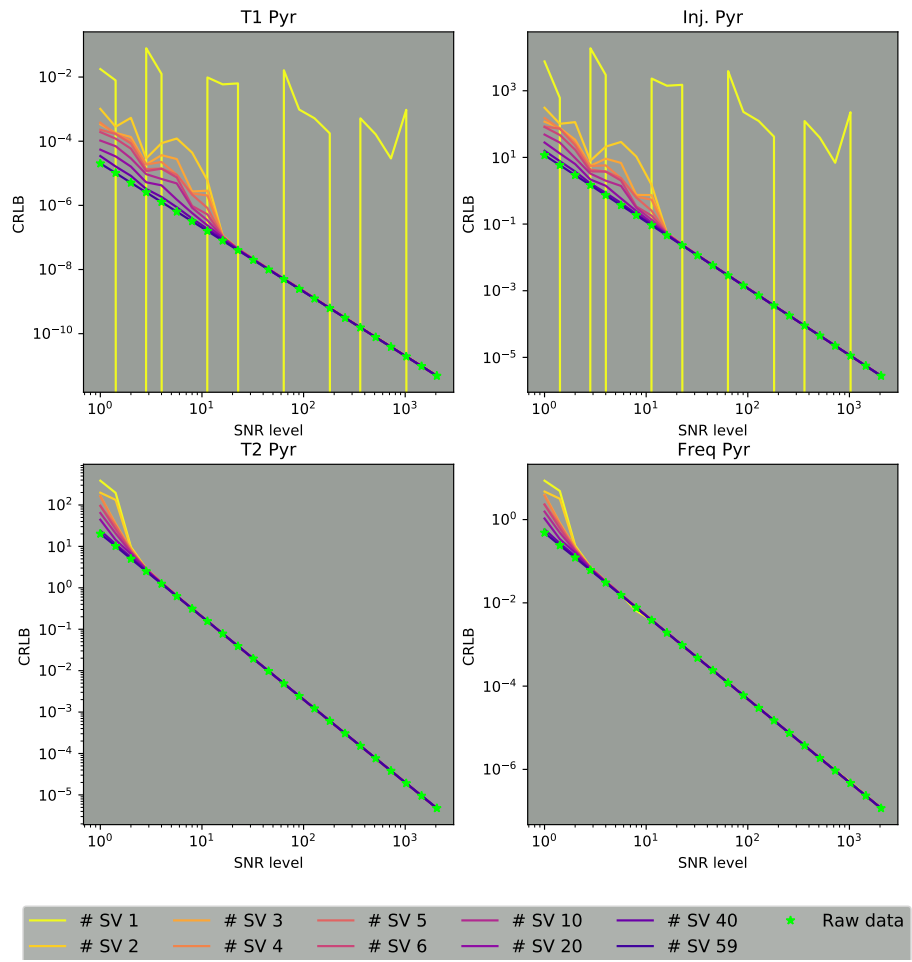

Figure 17: The CRLB for the pyruvate parameters. From the upper left corner in clockwise order: the effective $T_1$ relaxation rate, the injection rate, the relaxation time $T_2$, and the offset frequency

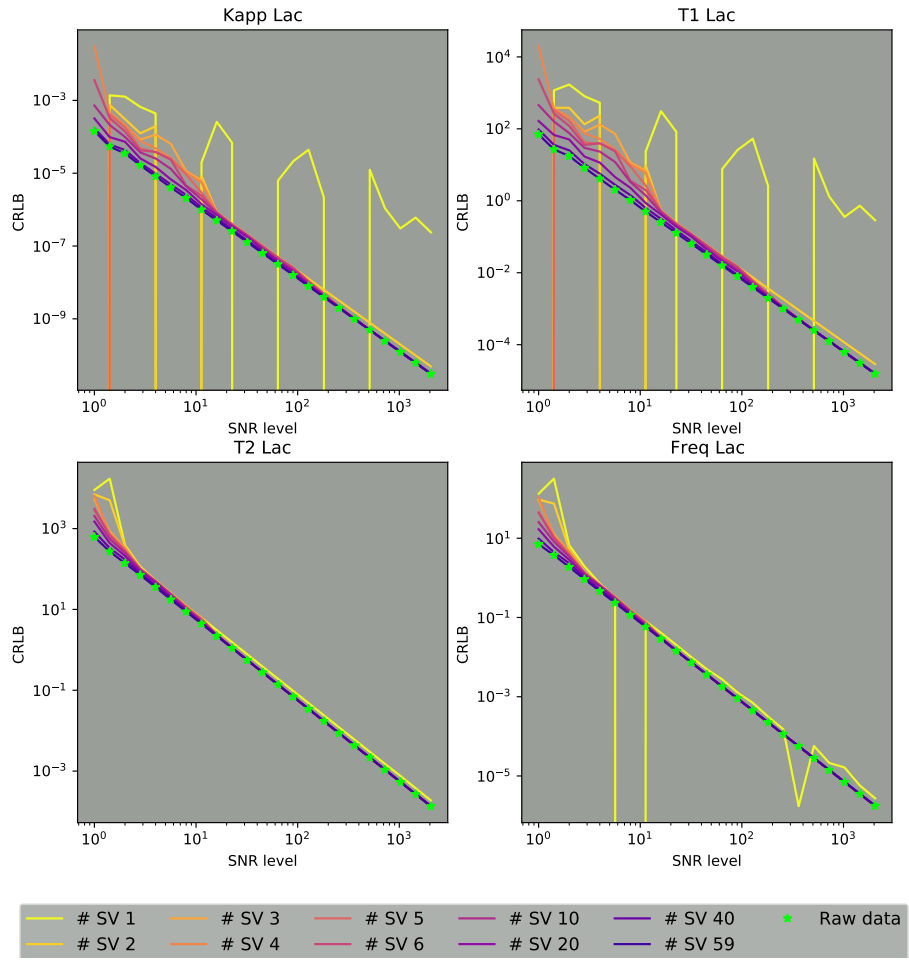

Figure 18: The CRLB for the lactate parameters. From the upper left corner in clockwise order: the effective reaction rate, the effective $T_1$ relaxation rate, the relaxation time $T_2$, and the offset frequency

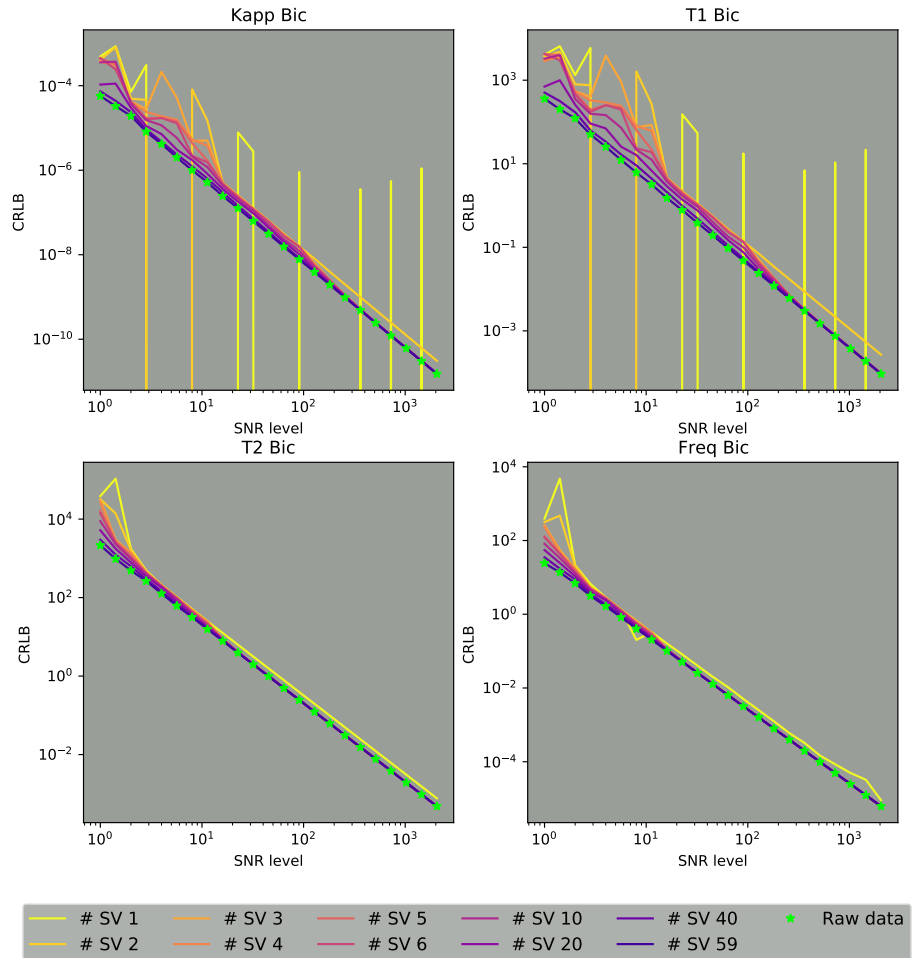

Figure 19: The CRLB for the bicarbonate parameters. From the upper left corner in clockwise order: the effective reaction rate, the effective $T_1$ relaxation rate, the relaxation time $T_2$, and the offset frequency

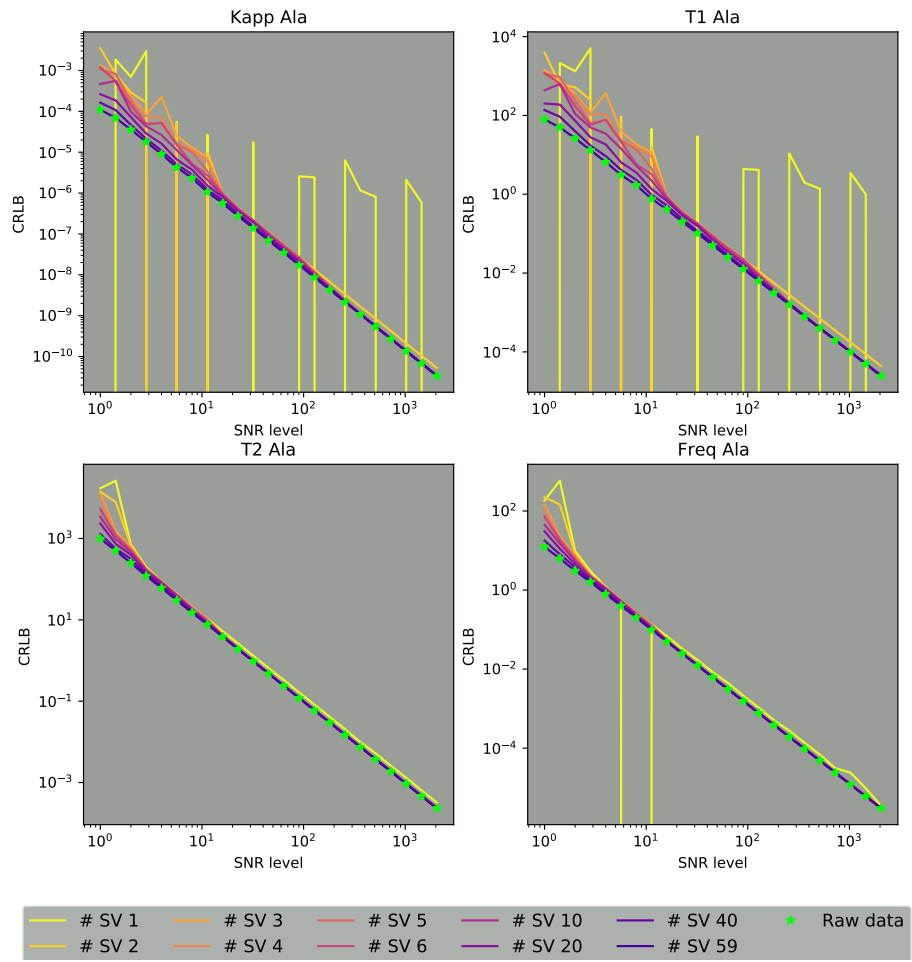

Figure 20: The CRLB for the alanine parameters. From the upper left corner in clockwise order: the effective reaction rate, the effective $T_1$ relaxation rate, the relaxation time $T_2$, and the offset frequency

# 5 Qualitative description of low-rank approximation filters

This entire section is a verbatim quote from the Appendix B of the Ph.D. thesis "Development of new experimental and data processing methods at critical signal-to-noise conditions in nuclear magnetic resonance" from page 101 to page 113.

## 5.1 System parameters for Montecarlo simulation

$$\text{FID}(t,T) = \sum_k M_k(T) e^{i2\pi f_k t} e^{\frac{-t}{T_{2k}}} \qquad k = [\text{pyr, ala, lac, bic}]. \quad (16)$$

Where $\text{FID}(t,T)$ is the value of the Free Induction Decay (FID) at the acquisition time $t$ after $T$ second from the injection, $M_k$ is the longitudinal polarization of the $k$ molecules, and $f_k$ is the molecules off-set frequency. The values of $M(T)_k$ is given by:

$$M_{pyr}(T) = \begin{cases} r_{inj} T_{1pyr} \left(1 - \exp\frac{t_{arr} - T}{T_{1pyr}}\right) & \text{if } t_{arr} \leq T < t_{end} \\ M_{pyr}(t_{end}) \exp\frac{t_{end} - T}{T_{1pyr}} & T \geq t_{end}, \end{cases} \quad (17)$$

where $r_{inj}$ is the injection rate, $t_{arr}$ and $t_{end}$ are respectively the starting time and the end time of injection. While for the other metabolite we have:

$$M_l(T) = \begin{cases} A_l \left(T_{1l}\left(1 - e^{\hat{T}_l}\right) - T_{1pyr}\left(1 - e^{\hat{T}_{pyr}}\right)\right) & \text{if } t_{arr} \leq T < t_{end} \\ B_l \left(e^{\hat{T}_l} - e^{\hat{T}_{pyr}}\right) + M_l(t_{end}) e^{\hat{T}_l} & T \geq t_{end} \end{cases} \quad (18)$$

$$\hat{T}_l = \frac{t_{arr} - T}{T_{1l}} \qquad\qquad \hat{T}_{pyr} = \frac{t_{arr} - T}{T_{1pyr}} \quad (19)$$

$$A_l = \frac{r_{inj} k_l}{1/T_{1pyr} - 1/T_{1l}} \qquad\qquad B_l = \frac{M_l(t_{end}) k_l}{1/T_{1pyr} - 1/T_{1l}} \quad (20)$$

Equation 16, 17 and 18 describe the model used for the Montecarlo simulation, the numerical values for the parameter used is reported in table 1.

| Metabolite | Inj. Rate ($s^{-1}$) | $T_{1eff}$ (s) | $r_{eff}$ ($s^{-1}$) | $T_2$ (ms) | Freq (Hz) |
|---|---|---|---|---|---|
| Pyruvate | 2.354e-2 | 19.998 | - | 49.996 | 2.158e-06 |
| Lactate | - | 13.178 | 1.899e-2 | 50.026 | -392.000 |
| Bicarbonate | - | 18.472 | 7.688e-3 | 50.061 | 321.309 |
| Alanine | - | 12.578 | 1.498e-2 | 50.031 | -270.000 |

Table 1: Model parameters are used for the Montecarlo simulations, rounded to the third significant digit

## 5.2 Qualitative description of low-rank approximation filters

The dataset under study has two dimensions[2]: the first one describes how the signal evolves during the Nuclear Magnetic Resonance (NMR) spectroscopy measurement and in particular how the transverse magnetization changes after the radio-frequency pulses; the second one describes how the longitudinal magnetization evolves between the NMR spectroscopy experiments. The metabolite could be identified using the information on their Larmor frequency contained in the first dimension of the dataset. Once the metabolites have been identified, the second dimension could be studied to identify the apparent conversion rate for each metabolite. The first dimension could be called the spectral dimension, while the second one could be called the metabolic dimension. Each dimension could be decomposed into some base components. This section compares the data components naturally arising from the metabolic conversion of pyruvate model with the components obtained using the SVD in a data drive approach. The first kind of components, here called natural components[3], are identified considering the metabolic conversion model, each metabolite spectrum is a natural component for the spectral dimension, while the time evolution curves for the longitudinal magnetization for each metabolite is a natural component for the metabolic dimension.

### 5.2.1 Natural components

The natural components of this system were already described in the main body of this thesis, here we just report their shape in figure 21 to ease the comparison with the data driven components. Their values are obtained from equations 16, 17, and 18.

---

[2]The data acquired during the experiment could be arranged into a matrix
[3]Please note that this naming convention is informal and personal; this naming choice does not convey any deep insight on the properties of these components

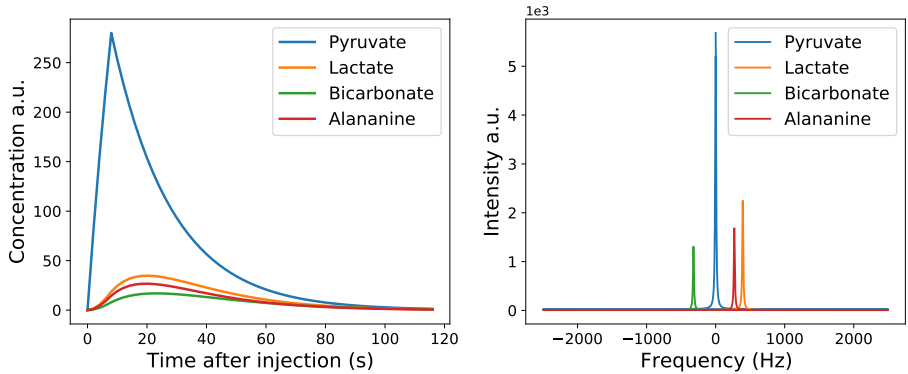

Figure 21: Left: time evolution of the metabolites' magnetization during the Montecarlo simulation. The curves were generated according to equations 17 and 18 . On the abscissa the time after injection in seconds, while on the ordinate the metabolites' concentrations in arbitrary units. Right: the simulation of the real part of the spectral signal 40 seconds after the injection, on the abscissa the frequency in Hz respect to the pyruvate Larmor frequency, on the ordinate the signal intensity in arbitrary units

### 5.2.2 Data driven components: the noiseless case

In the data driven approach the dataset components are obtained form the direct analysis of the acquired data and not from the study of the system's model. The SVD is used to construct the components from the dataset producing three matrices such as:

$$\mathbf{M} = \mathbf{U}\mathbf{S}\mathbf{V}^{\mathbf{H}} \qquad (21)$$

were $\mathbf{M}$ is a 59×2048 matrix containing the simulated data, $\mathbf{U}$ is a 59×59 unitary matrix which contains the information on the metabolic dimension of the dataset, $\mathbf{S}$ is 59×59 a diagonal matrix that contains the information regarding the weights of the components and $\mathbf{V}$ is a 2048×59 unitary matrix that contains the information on the spectral dimension of the dataset. The original data can be reconstructed from the sum and the product of the components. For example the $10^{th}$ spectrum is obtained by first multiplying the spectral components with their respective singular values (In this system we can consider just the first four components without a significant loss of accuracy). The singular values measure the contribution of the respective spectral components. Then, the weighted spectral components are multiplied with the elements of the $10^{th}$ metabolic component, now each spectral components is weighted for its general importance (the singular value) and its specific importance into the representation of the $10^{th}$ spectrum (the metabolic component elements).

The singular values intensity is reported in figure 24, only the first 4 singular values are different from zero considering the machine precision. Therefore, in the following analysis we will consider only the elements of **U** and **V** corresponding to these four singular values.

Figure 22 reports the first four components of the spectral dimension obtained by SVD analysis. We reported in blue the real values of the spectra and in orange the imaginary part. These components are quite different from the natural one, each component contains information on all the metabolites peaks mixed in different ways. Furthermore, the peaks "lineshape" is clearly non-Lorentzian and does not have the proper zero order phase factor between them. Indeed, these are not proper lineshapes originated form the physical nature of the system interaction, but just shapes needed to assure the linear independence of the components. Since the natural components are not orthogonal, the SVD derived components can not share their shape.

Figure 23 reports the first four components of the metabolic dimension obtained by SVD analysis. We reported in blue the real values of the spectra and in orange the imaginary part. The metabolic dimension's components share the same characteristic of the spectral dimension one. The discrepancy between the natural and the data driven components is exacerbated by the higher degree of non-orthogonality in the metabolic natural components compared to the spectral natural components[4]. Therefore, data driven metabolic components are even more different from natural components than spectral components. For example, the data-driven metabolic components are expressed as complex numbers while the natural components are real values as they describe the magnitude of polarization. To sum up, both data drive components are clearly different from the natural components of the previous subsection. While the natural components has a clear physical interpretation, these components does not. For example, the metabolic components are complex numbers which is inconsistent with the physical interpretation of metabolic components as a direct description of the longitudinal magnetization magnitude associated with a single metabolite. The discrepancy between the two sets of components derives from the mathematical properties of the SVD. Indeed, the matrices **U** and **V** should be unitary, therefore, their rows and columns (that constitute the data driven components) should be linear independent[5] while the natural components are linear dependent. Nevertheless, the direct interpretation of the data driven components is not needed for the creation of a low-rank approximation filter. Indeed, the objective of the SVD is to separate the signal form the noise and not to decompose the signals into a model explaining components set.

---

[4]The degree of non orthogonality could be better quantified taking the scalar product between the natural components after their normalization. A result of zero means that the components are orthogonal while a results of $\pm 1$ means that the components are collinear

[5]A set of vectors are linearly independent when each vector can not be written as the weighted sum of the remaining vectors

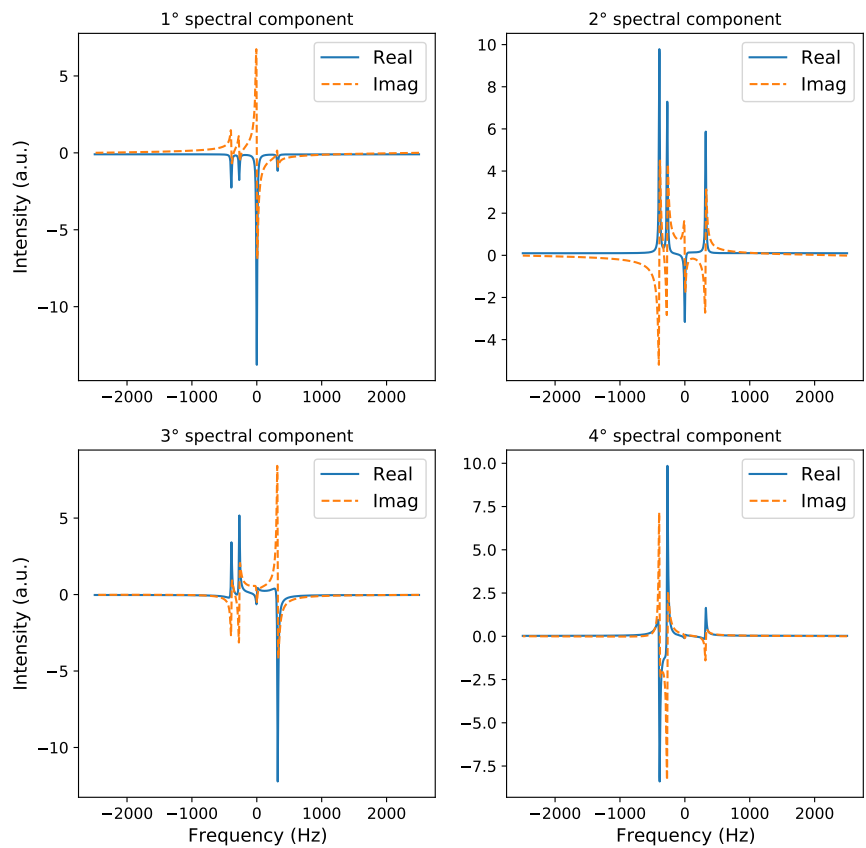

Figure 22: The first four components of the spectral dimension for the noiseless dataset, in blue the real value of the spectrum, while in orange the imaginary part. On the abscissa the frequency respect to the pyruvate Larmor frequency in Hz, on the ordinate the signal intensity in arbitrary units

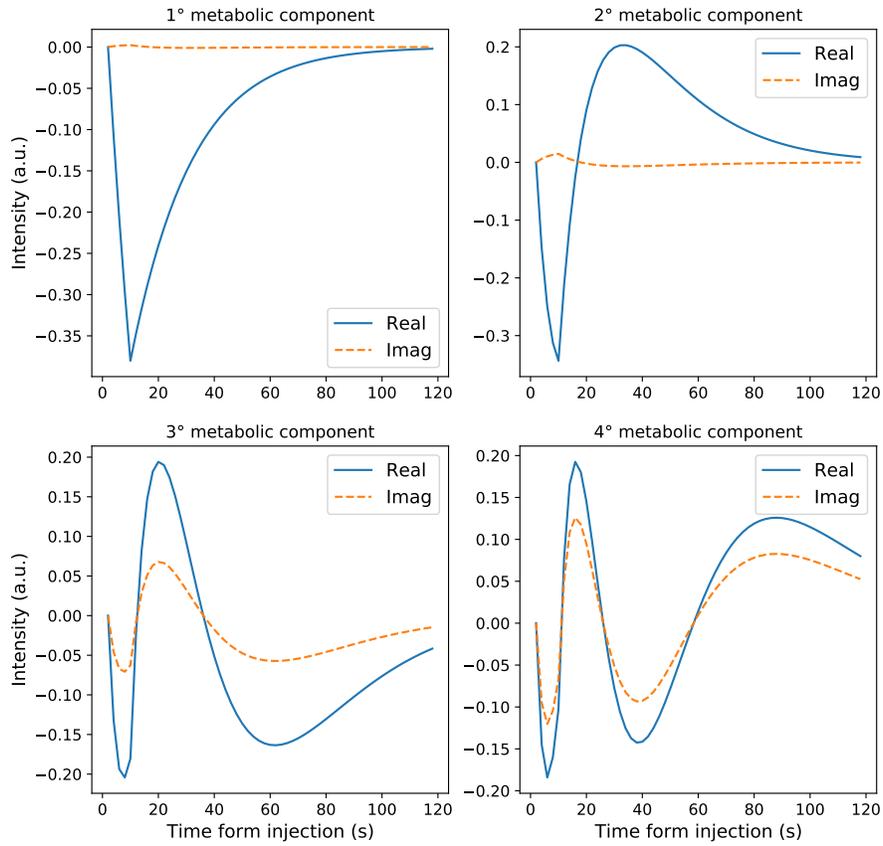

Figure 23: The first four components of the metabolic dimension for the noiseless dataset: in blue the real value of the longitudinal magnetization while in orange the imaginary part. On the abscissa the time after the injection in seconds, on the ordinate the signal intensity in arbitrary units

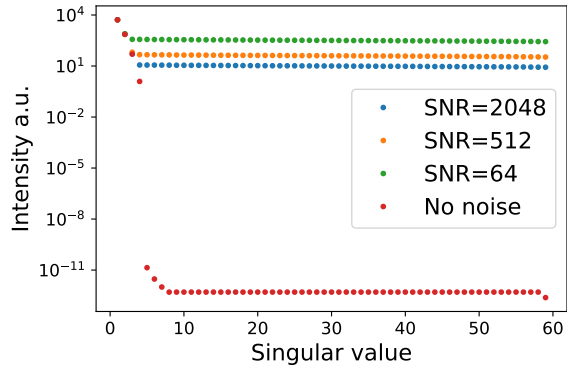

Figure 24: The singular values intensity obtained for different levels of noise added to the dataset. In blue, orange and green, respectively the curves obtained for the dataset with an SNR of 2048, 512 and 64. In red the curve obtained for the noiseless dataset. On the ordinate the singular value intensity depicted in logarithmic scale.

### 5.2.3 Data driven components: the noisy case

The presence of noise into the dataset corrupts the components obtained using the SVD. In this context the components are corrupted if they represent not only the true signal but also a significant fraction of the noise. This intuitive definition tries to capture a quite complex mathematical phenomenon. The SVD decomposition is not aware of the signal and noise nature, the SVD is just an instrument to solve the low-rank approximation problem, i.e. to find the best approximation to the original dataset using a fixed number of linearly independent components. Since noise and signal are not linearly independent the SVD can not perfectly separate the noise from the signal and part of the signal gets corrupted by the noise because the addition of a noise portion reduces the overall distance between the approximated dataset and the original noisy dataset.

Figure 25 represents the effect of four different noise levels on the first four components of the spectral dimension. The blue, orange and green curves are, respectively, obtained form a dataset with a SNR of 2048, 512 and 64. The red curves are obtained from a noiseless dataset. For ease of representation we reported just the real part of the components: it should be notate that part of the differences

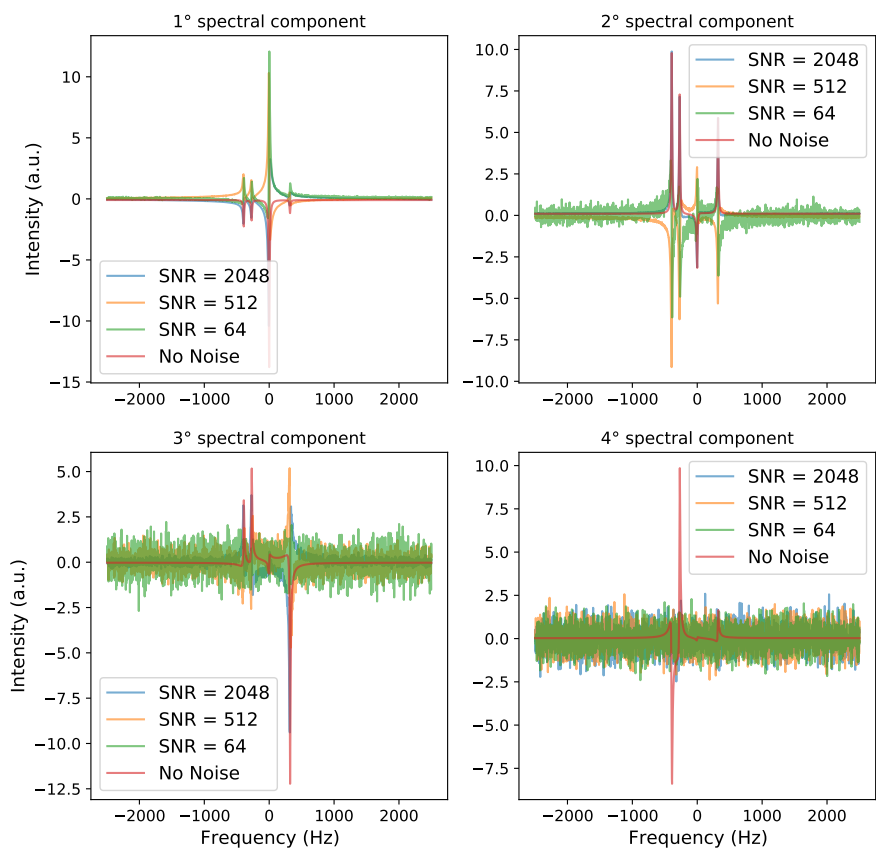

Figure 25: The real part of the first four components of the spectral dimension in the presence of four different noise levels are represented in this figure. The blue, orange and green curves are, respectively, obtained form a dataset with a SNR of 2048, 512 and 64. The red curves are obtained from a noiseless dataset. On the abscissa the frequency respect to the pyruvate Larmor frequency in Hz, on the ordinate the signal intensity in arbitrary units

between the curves is due to the presence of a zero order phase factor due to the SVD properties. Therefore, the smoothness of the curves is a better parameter to monitor the noise-related corruption for the data driven components compared to the curve magnitude. Indeed, the pure signal components are smooth because the NMR spectrum is a continuous function of the frequency, while the noise is a Gaussian random process that creates a rough curve. Figure 26 represents the effect of four different noise levels on the first four components of the metabolic dimension using the same convention employed in figure 25. Since both figures depict the components' corruption due to the noise, we can focus our attention just on figure 26, which is easier to analyze. The same observations hold for figure 25.

We can summarize the results of our visual inspection of the curve smoothness as:

- The first component is smooth for all the SNR considered.
- The second component is rough just for the lowest SNR level considered, 64.
- Only the highest SNR level has an almost smooth third component.
- The fourth component is rough for all the SNR levels considered.

All these results are in perfect accord with the analysis of the singular values intensity curve in figure 24. A singular value is usable in the signal reconstruction if it is before the knee in the curves. Therefore, there are 4 singular values usable for the dataset with SNR equal to 2048, and 3 for the other two datasets.

## 5.3 Qualitative evaluation of low-rank filter performance

The qualitative evaluation of the low-rank filter performance is carried out with a visual inspection of few spectra at different SNR levels. This is a qualitative analysis in the sense that it does not provide quantitative information on the underling physical system. The goal of hyperpolarized metabolic tracer experiment is to quantify and study the effective reaction rate. Therefore, the efficacy of a filter should be evaluated on its ability to improve the estimation of the effective reaction rate. The direct inspection of the spectra or the evaluation of the mean square errors between the filtered and the pure signals do not provide clear information on the ability of the filter to improve the estimation of the effective reaction rate, or other model parameters.

In figure 27 the effect of the low-rank filter on the spectrum at different SNR levels 4,8,16, and 32 is reported. The SNR level is calculated on the highest intensity signal of the experiment, the signal acquired at the end of the injection, 14 second after the beginning of the injection. I chose this SNR levels because they provide a clear insight on the efficacy of the methods for retrieving clear spectra even in the presence of low SNR. While this is not a sufficient condition to improve the estimation of the parameter describing the system reaction it is still a notable result and can be useful for analysis based on peak integration. To further highlight the effect of the low-rank filter the spectra after 20 and 50 second are reported since both this times feature a lower SNR compared to the maximum of each experiment and also have lower differences in the peaks' intensity compared to the signal after 14 seconds. The spectrum simulating the signal acquired after 20 and 50 seconds from injection are reported, respectively, on the left and the right column of the figure panel. The noisy signal is reported with light blue dots, the SVD filtered signal is reported with a orange solid line and the pure signal is reported with a red solid line. The SVD filter was constructed using the first 3 singular values based on the inspection of the singular values intensity in figure 24.

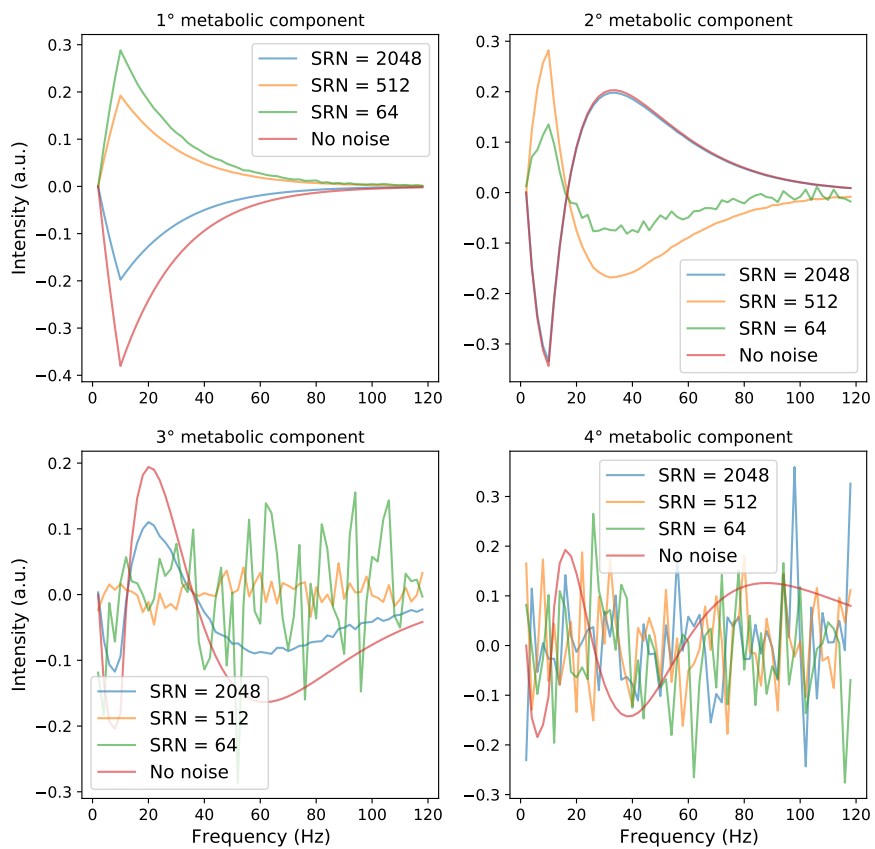

Figure 26: The real part of the first four components of the metabolic dimension in the presence of four different noise levels are represented in this figure. The blue, orange and green curves are, respectively, obtained form a dataset with a SNR of 2048, 512 and 64. On the abscissa the time after the injection in seconds, on the ordinate the signal intensity in arbitrary units

The most interesting results are obtained for the 50 seconds after injection spectra, the SVD filter allows for a clear identification of the peaks and their area even when their intensity is lower than noise level, for example in the first two rows of the figure panel. This result is particularly interesting for the pyruvate metabolites, their lower intensity makes their identification harder and the estimation of the peaks area is more susceptible to the presence of noise compared with the stronger pyruvate peak. While these noise reduction performances do not directly translate into an increased quality of the metabolic parameters estimated from the filtered dataset, they are still useful for the qualitative inspection of the spectra.

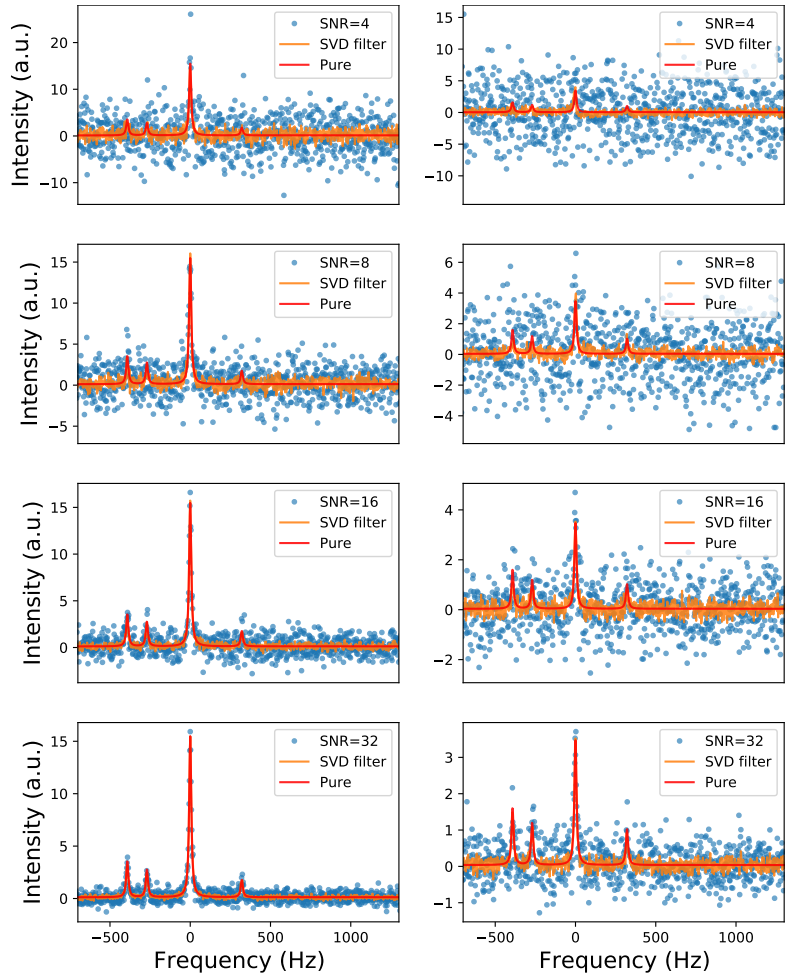

Figure 27: Effect of the low-rank filter on spectra simulated at different times after injection. Left: simulation of signal acquired after 20 seconds. Right: simulation of signal acquired after 50 seconds. The noisy signal is reported with light blue dots, the SVD filtered signal is reported with a orange solid line and the pure signal is reported with a red solid line.